\documentclass[aps,prfluids,onecolumn,superscriptaddress]{revtex4-2}
\usepackage[pdfauthor={R. Hartmann, R. Verzicco, D. Lohse, R.J.A.M. Stevens}, pdftitle={SHEAROTpaper}, bookmarks=true, colorlinks=true, linkcolor=RoyalBlue, urlcolor=blue, citecolor=OliveGreen, pdftex, bookmarks=true, hyperindex=true]{hyperref}

\usepackage{graphicx}

\usepackage{amsmath}
\usepackage{amssymb}

\usepackage{color}
\usepackage[dvipsnames]{xcolor}

\renewcommand{\Pr}{\mathrm{Pr}}
\newcommand{\Ra}{\mathrm{Ra}}
\newcommand{\Ek}{\mathrm{Ek}}
\newcommand{\Ro}{\mathrm{Ro}}
\newcommand{\Fr}{\mathrm{Fr}}
\newcommand{\Gr}{\mathrm{Gr}}
\newcommand{\iEk}{\mathrm{Ek}^{-1}}
\newcommand{\iRo}{\mathrm{Ro}^{-1}}
\newcommand{\iEkopt}{\mathrm{Ek}^{-1}_\mathrm{opt}}
\newcommand{\iRoopt}{\mathrm{Ro}^{-1}_\mathrm{opt}}

\newcommand{\Nu}{\mathrm{Nu}}

\begin{document}

\title{Optimal heat transport in rotating Rayleigh-B\'{e}nard convection at large Rayleigh numbers}

\author{Robert Hartmann}
\email[]{r.hartmann@utwente.nl}
\affiliation{Physics of Fluids Group and Twente Max Planck Center, J.M. Burgers Center for Fluid Dynamics, University of Twente, 7500 AE Enschede, The Netherlands}

\author{Guru S. Yerragolam}
\affiliation{Physics of Fluids Group and Twente Max Planck Center, J.M. Burgers Center for Fluid Dynamics, University of Twente, 7500 AE Enschede, The Netherlands}

\author{Roberto Verzicco}
\affiliation{Physics of Fluids Group and Twente Max Planck Center, J.M. Burgers Center for Fluid Dynamics, University of Twente, 7500 AE Enschede, The Netherlands}
\affiliation{Dipartimento di Ingegneria Industriale, University of Rome 'Tor Vergata', 00133 Rome, Italy}
\affiliation{Gran Sasso Science Institute, 67100 L'Aquila, Italy}

\author{Detlef Lohse}
\affiliation{Physics of Fluids Group and Twente Max Planck Center, J.M. Burgers Center for Fluid Dynamics, University of Twente, 7500 AE Enschede, The Netherlands}
\affiliation{Max Planck Institute for Dynamics and Self-Organisation, 37077 G\"ottingen, Germany}

\author{Richard J.A.M. Stevens}
\affiliation{Physics of Fluids Group and Twente Max Planck Center, J.M. Burgers Center for Fluid Dynamics, University of Twente, 7500 AE Enschede, The Netherlands}

\date{August 28, 2023}

\begin{abstract}
The heat transport in rotating Rayleigh-B\'{e}nard convection (RBC) can be significantly enhanced for moderate rotation, i.e., for an intermediate range of Rossby numbers $\Ro$, compared to the non-rotating case. At Rayleigh numbers $\Ra\lesssim 5\cdot10^8$, the largest enhancement is achieved when the thicknesses of kinetic and thermal boundary layer are equal. However, experimental and numerical observations show that, at larger $\Ra$ ($\gtrsim 5\cdot10^8$), the maximal heat transport starts to deviate from the expected optimal boundary layer ratio and its enhancement amplitude decreases drastically. We present data from direct numerical simulations of rotating RBC in a periodic domain in the range of $10^7\leq\Ra\leq10^{10}$ and $0\leq\iRo\leq40$ for Prandtl number $\Pr=4.38$ and $6.4$ (corresponding to Ekman numbers $\Ek\gtrsim10^{-6}$) to identify the reason for the transition to this large-$\Ra$ regime of heat transport enhancement. Our analysis reveals that the transition occurs once the bulk flow at the optimal boundary layer ratio changes to geostrophic turbulence for large $\Ra$. In that flow state, the vertically coherent vortices, which support heat transport enhancement by Ekman pumping at smaller $\Ra$, dissolve into vertically decorrelated structures in the bulk such that the enhancing effect of Ekman pumping and the influence of the boundary layer ratio become small. Additionally, more heat leaks out of the Ekman vortices as the fraction of thermal dissipation in the bulk increases. We find that the rotation-induced shearing at the plates helps to increase the thermal dissipation in the bulk and thus acts as a limiting factor for the heat transport enhancement at large $\Ra$: beyond a certain ratio of wall shear stress to vortex strength, the heat transport decreases irrespectively of the boundary layer ratio. This $\Pr$-dependent threshold, which roughly corresponds to a bulk accounting for $\approx1/3$ of the total thermal dissipation, indicates the maximal heat transport enhancement and the optimal rotation rate $\iRoopt$ at large $\Ra$.
\end{abstract}

\maketitle

\section{\label{sec:intro}Introduction}

Most turbulent convection phenomena, which surround us in nature, are inaccessible for both experiments and numerical simulations due to their extreme parameters. For such systems, effective scaling laws have become a handy tool to relate the flow dynamics within the accessible range of control parameters to those of the system in nature, e.g., the convection in planetary cores \citep{aurnou_rotating_2015,cheng_laboratory-numerical_2015} or subsurface oceans of Jovian and Saturnian moons \citep{soderlund_ocean_2019}. However, it requires that the underlying physics does not fundamentally change in order for the effective scaling relations to remain valid. A very prominent example, for which this implicit assumption fails, is the phenomenon of enhanced heat transport in rotating Rayleigh-B\'{e}nard convection (RBC) \citep[e.g.,][]{chandrasekhar_instability_1953,rossby_study_1969,stevens_heat_2013,kunnen_geostrophic_2021,ecke_turbulent_2023}. In this model system of thermal buoyancy-driven convection altered by rotation, one can observe a significantly enhanced heat transport compared to the non-rotating case \citep[e.g.,][]{rossby_study_1969,julien_rapidly_1996,kunnen_heat_2006,liu_heat_2009}. The optimal rate of rotation, which is required to maximize the heat transport, follows a prescribed scaling behavior when thermal driving is relatively small \citep{king_heat_2012,yang_what_2020}. However, when thermal driving becomes larger, the enhancement vanishes and its rotational optimum deviates from the predicted scaling \citep{yang_what_2020}. This paper further investigates the reason(s) behind the vanishing heat transport enhancement in rotating RBC towards large thermal driving, building on the data of \citet{yang_what_2020}. Thereby, we reveal new insights in the transitions of the underlying flow dynamics, which help to put the phenomenon of rotation-induced heat transport enhancement into the larger picture of the transition from non-rotating to rapidly-rotating RBC \citep{kunnen_geostrophic_2021}.\\

The canonical rotating RBC system is controlled by three dimensionless parameters: the Prandtl number $\Pr$ describing the fluid properties, the Rayleigh number $\Ra$ setting the strength of thermal driving, and the inverse Rossby number $\iRo$ as a measure of the rotation rate $\Omega$:
\begin{equation}
\Pr=\frac{\nu}{\kappa}
\hspace{0.2cm}\text{,}\hspace{0.2cm}
\Ra=\frac{\alpha g \Delta T H^3}{\nu\kappa}
\hspace{0.2cm}\text{,}\hspace{0.2cm}
\iRo=\frac{2\Omega H}{\sqrt{\alpha g \Delta T H}}
\hspace{0.2cm}\text{.}
\end{equation}
Therein, $\nu$ is the kinematic viscosity, $\kappa$ the thermal diffusivity, $\alpha$ the isobaric thermal expansion coefficient, $g$ the gravitational acceleration, and $\Delta T$ and $H$ the temperature difference and distance between upper and lower plate, respectively. Alternatively, the influence of rotation and buoyancy can be expressed in terms of the Ekman number $\Ek=\Ro\sqrt{\Pr/\Ra}$ and the supercriticality $\widetilde{\Ra}=\Ra/\Ra_c$, $\Ra_c=8.7\,\Ek^{-4/3}$ denoting the critical Rayleigh number required to initiate convection under strong rotation \citep{chandrasekhar_hydrodynamic_1961}. The key response parameter is the heat transport, which, in the dimensionless form, is given by the Nusselt number $\Nu=Q H/(\kappa\Delta T)$, where $Q$ is the heat flux from the bottom to the top plate.\\

\begin{figure}
\centering
\includegraphics[width=\textwidth]{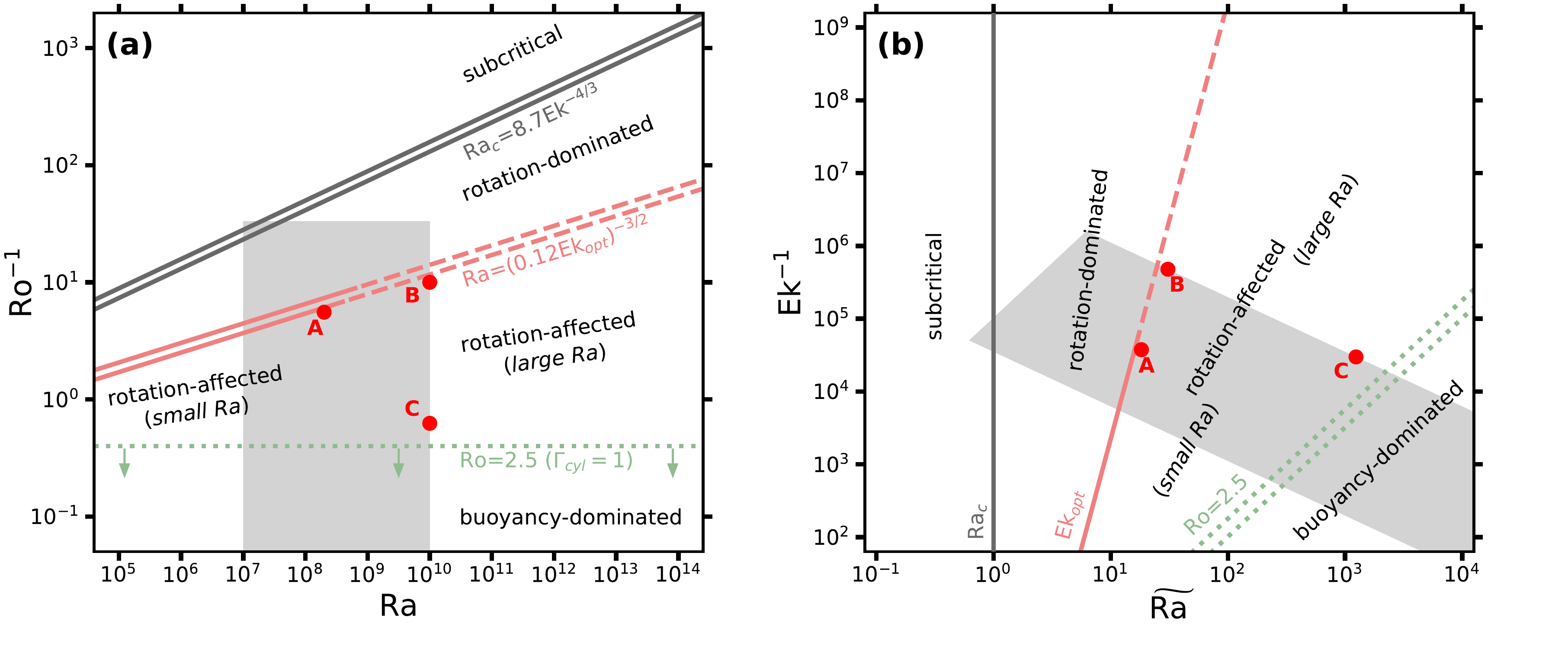}\\
\vspace{-0.3cm}
\caption{\label{fig:RegMap}Schematic regime diagram of rotating RBC in the parameter space of $\Ra$ and $\iRo$ \textbf{(a)} or $\widetilde{\Ra}$ and $\iEk$ \textbf{(b)}. The lines mark the basic regime transitions: the onset of convection based on the critical Rayleigh number $\Ra_c$ \citep{chandrasekhar_hydrodynamic_1961} (solid gray), the boundary layer crossing ($\lambda_\Theta/\lambda_u\approx1$) corresponding to the maximal heat transport enhancement $\Nu_\text{max}/\Nu_0$ at small $\Ra$ (solid red) \citep{yang_what_2020} being extrapolated to large $\Ra$ (dashed red), and the onset of heat transport enhancement in a $\Gamma=1$ cylinder \citep{weiss_finite-size_2010}. The latter is only meant as a guide to the eye, as it shifts to smaller $\iRo$ for increasing $\Gamma$ and appears gradual in the periodic setup. Double lines indicate the $\Pr$ dependence of the transitions in one or the other representation with respect to the used values $\Pr=4.38$ and $6.4$. The shaded area shows the parameter range of the DNSs in this study. A, B, and C mark exemplary cases that are discussed in detail later in this paper.}
\end{figure}

Starting from the classical non-rotating Rayleigh-B\'{e}nard case ($\iRo=0$) with heat transport $\Nu_0$ for fixed values of $\Ra$ and $\Pr>1$, the system traverses three main regimes with increasing rotation \citep[e.g.,][]{kunnen_role_2011,kunnen_geostrophic_2021,ecke_turbulent_2023} (Fig.~\ref{fig:RegMap}). These regimes are characterized by different flow states, which also affect the heat transport behavior differently:\\

(i) When the system rotates only weakly, the system is in the buoyancy-dominated (also called rotation-unaffected) regime \citep[e.g.,][]{kunnen_geostrophic_2021,ecke_turbulent_2023}. Therein, the flow is basically unaffected by the Coriolis force and often sustains the typical large-scale convection rolls in the bulk \citep[e.g.,][]{brown_reorientation_2005,poel_pencil_2015}. Consequently, the heat transport remains unchanged ($\Nu\approx\Nu_0$).\\

(ii) When the strength of rotation reaches a moderate level, the system enters the rotation-affected regime. The transition between the two regimes is sharp in laterally confined RBC containers \citep[e.g.,][]{kunnen_breakdown_2008,weiss_finite-size_2010,rajaei_transitions_2016,alards_sharp_2019} but rather gradual in horizontally periodic domains \citep{kunnen_heat_2006,yang_what_2020}. In the rotation-affected regime, the influence of the Coriolis force starts to alter the flow. At relatively small $\Ra$, the large-scale circulation gets replaced by vertically coherent vortices \citep[e.g.,][]{julien_rapidly_1996,vorobieff_vortex_1998,kunnen_breakdown_2008}. Ekman pumping feeds these vortices with hot (cold) fluid from the bottom (top) boundary layer and thereby enhances the heat transport $\Nu>\Nu_0$, when $\Pr>1$ \citep[e.g.,][]{rossby_study_1969,kunnen_heat_2006,liu_heat_2009,weiss_heat-transport_2016}. The enhancement is most efficient and reaches its maximum $\Nu_\text{max}$ when the thicknesses of thermal and kinetic boundary layer are approximately equal $\lambda_\Theta/\lambda_u\approx1$ \citep{stevens_optimal_2010,yang_what_2020}. At sufficiently large $\Ra$, the system experiences a rather turbulent flow state of (geostrophically) \textit{unbalanced boundary layers} \citep{cheng_heuristic_2018} with much less coherent structures in the bulk \citep{aguirre-guzman_force_2021}. In this case, no significant enhancement of the heat transport has been observed yet, though the rotation-affected regime for $\Ra\gg10^{10}$ and $\Pr>1$ is relatively unexplored \citep{cheng_laboratory-numerical_2015,cheng_laboratory_2020,lu_heat-transport_2021}. \citet{yang_what_2020} reported evidence for the transition between these two flow states in the rotation-affected regime to appear at $\Ra\approx5\cdot10^{8}$ for waterlike $\Pr$ numbers.\\

(iii) When the system is rotating rapidly, the flow transitions to the rotation-dominated regime. Therein, the heat transport $\Nu$ steeply decreases with increasing rotation until convection ends at $\Ek_c=(8.7/\Ra)^{3/4}$. Four characteristic flow morphologies have been observed in the rotation-dominated regime \citep[e.g.,][]{sprague_numerical_2006,julien_statistical_2012,nieves_statistical_2014,stellmach_approaching_2014,cheng_laboratory-numerical_2015}. In a small range above the onset of convection $1<\widetilde{\Ra}\lesssim2$, the flow assembles a non-turbulent cellular structure. For larger $\widetilde{\Ra}$, the flow again organizes in the aforementioned vertically coherent vortices. When these vortices bridge the entire bulk, they are typically called convective Taylor columns. When these vortices start to loose their vertical coherence and do not fully reach the opposite plate, the remaining vortical plumelike structures are often referred to as the \textit{plume state} of rotation-dominated RBC. At very large $\widetilde{\Ra}$, i.e., large $\Ra$ and small $\Ek$, the vertical coherence of the Ekman vortices gets fully lost. The flow is in so-called geostrophic turbulence, a flow state of decorrelated small-scale vortex structures in the bulk, which tend to assemble in large-scale vortices \citep[e.g.,][]{stellmach_approaching_2014,rubio_upscale_2014,aguirre-guzman_competition_2020,aguirre-guzman_force_2021}. The formation of large-scale vortices is, however, more prominent for stress-free plates and $\Pr<1$ \citep{stellmach_approaching_2014,kunnen_transition_2016}. For no-slip conditions at the plates and $\Pr\geq1$ either no such vortex could be observed \citep{kunnen_transition_2016} or only at extremely large $\widetilde{\Ra}$ \citep{aguirre-guzman_competition_2020}.\\ 

For small $\Ra$, the heat transport maximum at $\lambda_\Theta/\lambda_u\approx1$ is naturally seen as the transition between the rotation-affected and the rotation-dominated regimes \citep{king_boundary_2009,kunnen_role_2011,kunnen_geostrophic_2021}. Based on basic theoretical boundary layer arguments, \citet{king_heat_2012} derived a scaling law $\Ra\propto\Ek^{-3/2}$ (see also Sec.~\ref{subsec:BLratio}) that by a prefactor resembles at once (i) a constant boundary layer ratio, (ii) the rotation-affected to rotation-dominated transition, and (iii) the location of the heat transport maximum, i.e., its optimal rotation rate \citep{king_heat_2012,king_scaling_2013,yang_what_2020}. However, it is less intuitive to define an accurate transition between rotation-affected and rotation-dominated regimes at large $\Ra$, where the heat transport enhancement vanishes. Often, the term geostrophic regime is also used as a synonym for the rotation-dominated regime. However, the geostrophic regime is rather defined based on the predominant geostrophic force balance (between pressure and Coriolis force) in the flow. In some cases, the geostrophic balance can still be maintained in the rotation-affected regime \citep{kunnen_geostrophic_2021}. Thus, strictly speaking, the geostrophic regime encompasses a different part of the $(\Ra,\iRo,\Pr)$ parameter space than the rotation-dominant regime defined by the boundary layer crossing. Hence, there exist many different scaling approaches for regime or flow state transitions at large $\Ra$ \citep[e.g.,][]{julien_heat_2012,ecke_heat_2014}. We refer the reader to the review by \citet{kunnen_geostrophic_2021} for a detailed overview of these transitions. In this study, we stick to the aforementioned boundary layer crossing ($\lambda_\Theta/\lambda_u\approx1$) to define the rotation-affected to rotation-dominated transition over the entire range of $\Ra$.\\

The objective of this study is to identify what mechanisms affect heat transport enhancement at large $\Ra$, i.e., lead to the strong reduction of the magnitude of enhancement and control the optimal rate of rotation $\iRoopt$ of the heat transport maximum. After introducing the setup for our direct numerical simulations (DNSs) of rotating RBC (Sec.~\ref{sec:DNS}), we first present the heat transport enhancement at various $\Ra$ and work out the discrepancy between the observed $\iRoopt$ and the expected optimum based on boundary layer crossing at large $\Ra$ (Sec.~\ref{sec:HTEnBL}). In Sec.~\ref{sec:Flow}, we compare the flow morphology at small and large $\Ra$ to show that the onset of geostophic turbulence inhibits a largely enhanced heat transport at large $\Ra$. Finally, we identified what quantities correlate with the optimal rotation rate at large $\Ra$ and discuss why these might effectively control $\iRoopt$ (Sec.~\ref{sec:RaReg}).

\section{\label{sec:DNS}Governing equations \& numerical setup}

Rotating RBC is governed by the conservation of mass, momentum, and energy. Under Oberbeck-Boussinesq approximation, these conservation laws transform into the continuity equation, the Navier-Stokes equations, and the convection-diffusion equation of temperature, respectively. In dimensionless form, the equations are given with respect to the three control parameters $\Pr$, $\Ra$, and $\iRo$ as:
\begin{equation}
\begin{split}
\nabla\cdot\vec{u}=&0\hspace{0.2cm}\text{,}\\
\frac{\mathrm{d}\vec{u}}{\mathrm{d}t}=&-\nabla P +\sqrt{\frac{\Pr}{\Ra}}\nabla^2\vec{u}+\Theta\vec{e}_z-\frac{1}{\Ro}\vec{e}_z\times\vec{u}\hspace{0.2cm}\text{,}\\
\frac{\mathrm{d}\Theta}{\mathrm{d}t}=&\frac{1}{\sqrt{\Pr\,\Ra}}\nabla^2\Theta\hspace{0.2cm}\text{.}\\
\end{split}
\label{eq:NS}
\end{equation}
Therein, $\vec{u}$, $P$, and $\Theta$ are the normalized velocity, pressure, and temperature fields, respectively. The equations are normalized by the domain height $H$ and the free-fall velocity $U_0=\sqrt{\alpha g \Delta T H}$. The temperature is normalized as $\Theta=\frac{T-T_{\mathrm{top}}}{\Delta T}\in[0,1]$. The pressure field $P$ is reduced by the hydrostatic balance and centrifugal contributions.\\

Regarding Eqs.~\ref{eq:NS}, we consider Coriolis forcing from constant rotation around a vertical axis but neglect centrifugal buoyancy. This is equivalent to a rotational Froude number $\Fr=\Omega^2R/g=0$, where $R$ is the maximal radial distance from the rotation axis. The limit of $\Fr\ll1\to0$ is valid in most geo- and astrophysical contexts (e.g., $\Fr_{\mathrm{Earth}}\approx8.7\cdot10^{-5}$). Implicitly setting $\Fr=0$ allows to impose periodicity in the lateral directions. Hence, the simulations are performed in a Cartesian domain with isothermal plates at the top ($\Theta=0$) and the bottom ($\Theta=1$) and no-slip conditions for the velocity ($\vec{u}=0$), whereas velocity and temperature are periodic in both horizontal directions. The lateral extent of the domain is given by the variable width-to-height ratio $\Gamma=W/H$. In accordance with \citet{kunnen_transition_2016}, we ensure that $\Gamma\geq10\,l_c$, where $l_c=4.82\,\Ek^{1/3}$ is the most unstable wavelength of the convective instability \citep{chandrasekhar_hydrodynamic_1961,niiler_influence_1965,heard_asymptotic_1971}.\\

In this study, we cover the parameter space of $10^{7}\leq\Ra\leq10^{10}$ and $0\leq\iRo\leq40$ for two Prandtl numbers $\Pr=4.38$ and $\Pr=6.4$ (corresponding to $0\leq\iEk\leq1.6\cdot10^6$). Thereby, we draw on the data set from \citet{yang_what_2020} for our analysis and extend it with in total 81 newly conducted DNSs. The simulations solve the governing equations (Eqs.~\ref{eq:NS}) by a central second-order accurate finite-difference scheme based on a staggered grid discretization as presented in \citet{poel_pencil_2015} and \citet{ostilla-monico_multiple-resolution_2015}. The code has been often validated \citep[e.g.,][]{kooij_comparison_2018}. The size of the computational domain is $\Gamma N_z\times\Gamma N_z\times N_z$, where the number of grid points in the vertical direction $N_z$ increases with $\Ra$ up to $N_z=1024$ for $\Ra=10^{10}$. While the horizontal directions are uniformly spaced, a clipped Chebyshev-like clustering of grid points towards the plates is applied in the vertical direction. This ensures the full resolution of the Kolmogorov scales in the entire domain, as well as meeting the criteria for the resolution of the boundary layers given in \citet{shishkina_boundary_2010}. The dynamic time stepping in our simulations is controlled by a maximum CFL number and a maximum time step. The numerical parameters of the new simulations are summarized in App.~\ref{sec:appA} (Tabs.~\ref{tab:pr4},\ref{tab:pr6}). All newly conducted simulations are run extensively long and have reached a statistical convergence of less than $0.5\%$ for the integral flow quantities such as the Nusselt number $\Nu$.

\section{\label{sec:HTEnBL}Observed vs. expected heat transport enhancement}

\subsection{\label{subsec:HTE}Heat transport enhancement with increasing Rayleigh number}

\begin{figure*}
\centering
\includegraphics[width=\textwidth]{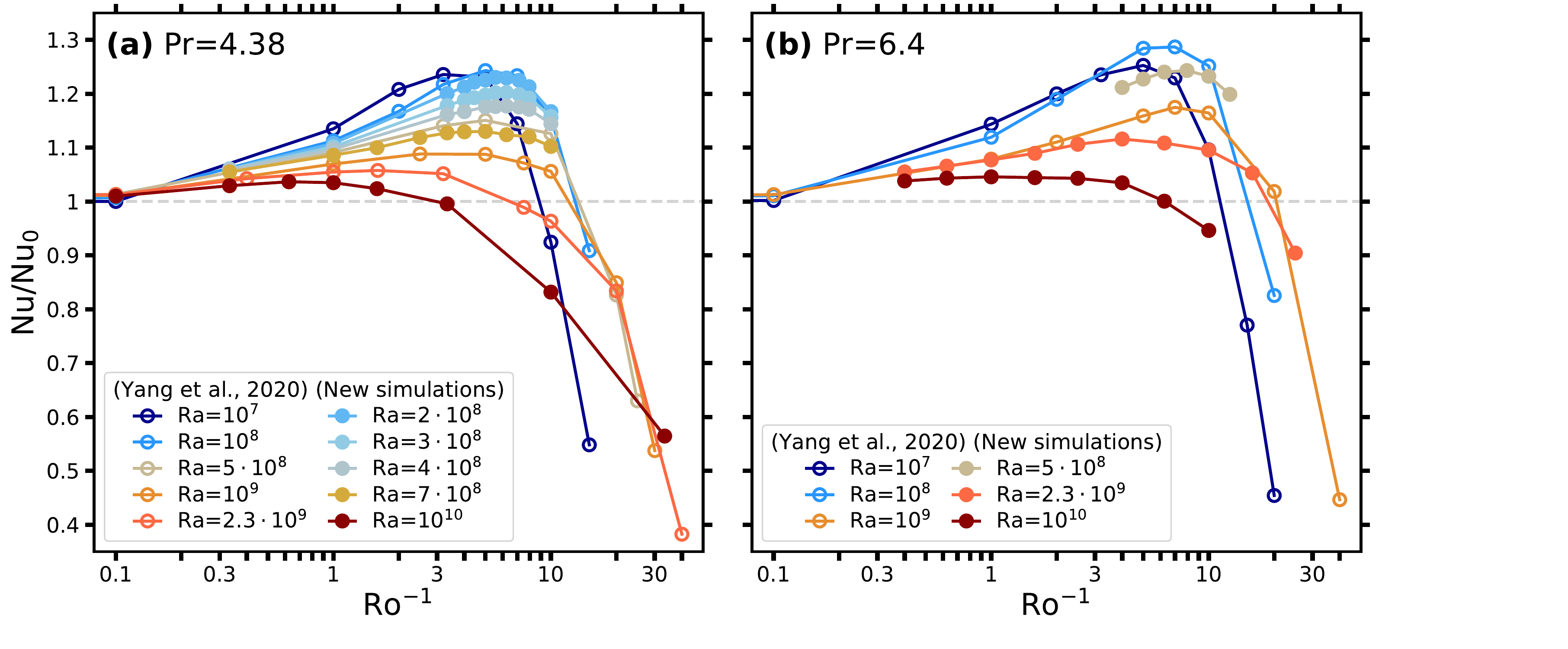}\\
\includegraphics[width=\textwidth]{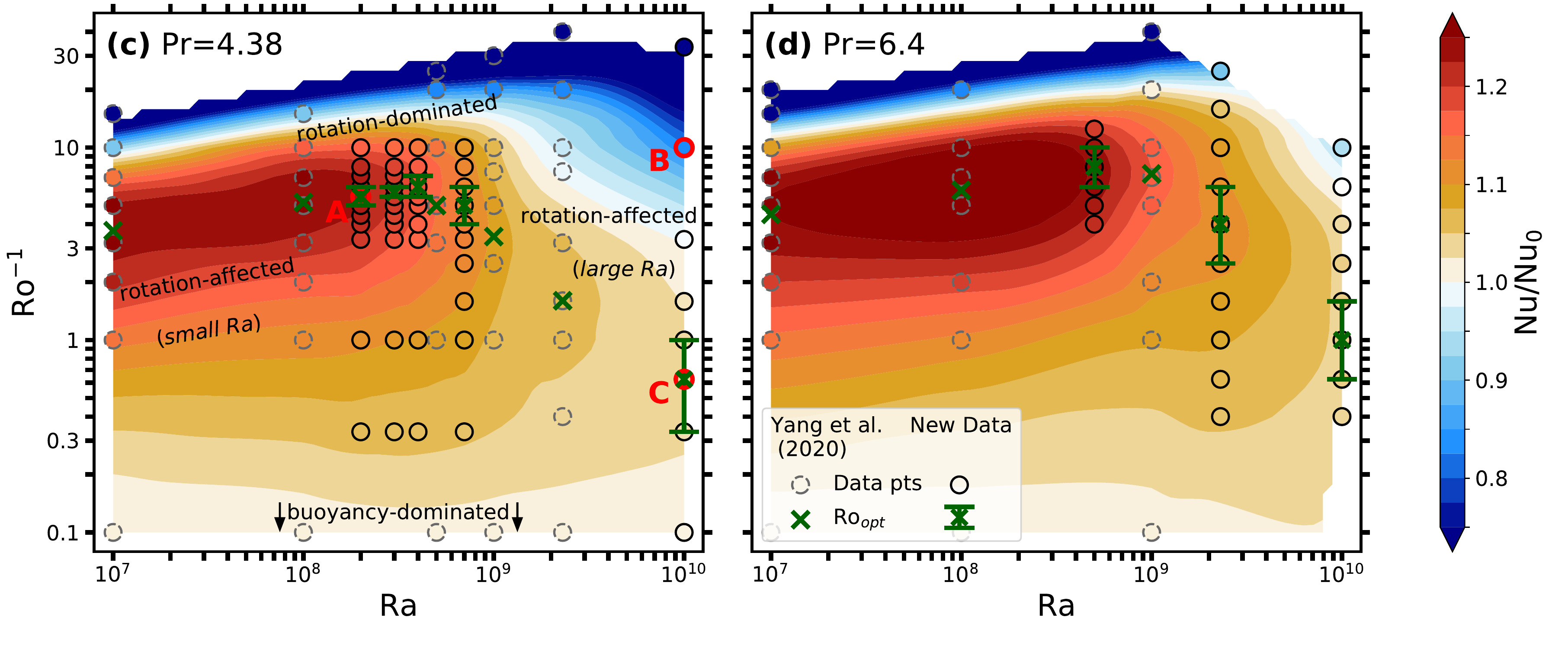}\\
\vspace{-0.3cm}
\caption{\label{fig:NuAvg}\textbf{(a,b)}~Normalized heat transport $\Nu/\Nu_0$ as function of the rotation rate $\iRo$ at various $\Ra$ for $\Pr=4.38$ and $\Pr=6.4$. Open symbols represent DNS data from \citet{yang_what_2020}, filled symbols newly conducted DNS data for this study. \textbf{(c,d)}~Normalized heat transport $\Nu/\Nu_0$ in the parameter space of rotation $\iRo$ and thermal driving $\Ra$ for $\Pr=4.38$ and $\Pr=6.4$ (dashed circles - data points from \citet{yang_what_2020}, closed circles - new data points, background - cubic interpolation). The crosses mark the maximal heat transport $\Nu_{\mathrm{max}}(\Ra)$ per $\Ra$. The error bars reflect the uncertainty of $\iRoopt$ due to the sampling of $\iRo$ cases, giving a conservative uncertainty estimate for the new data. In (c), the main flow regimes are indicated as orientation for the reader (see also Fig.~\ref{fig:RegMap}). A, B, and C mark the cases presented in Fig.~\ref{fig:FlowSnaps}.}
\end{figure*}

The response parameter of key interest is the dimensionless heat transport $\Nu$. In light of optimal rotation for heat transport enhancement, we naturally start by looking at the normalized heat transport $\Nu/\Nu_0$ as a function of the dimensionless rotation rate $\iRo$ for various fixed $\Ra$ per $\Pr$ (Fig.~\ref{fig:NuAvg}(a,b)). We compute the Nusselt number from the horizontal-temporal average $\left<\cdot\right>_\mathcal{H}=\left<\cdot\right>_{x,y,t}$ of the vertical temperature gradient at the plates: $\Nu=\left<-\partial_z\left<\Theta\right>_\mathcal{H}\right>_{z=\{0,1\}}$. For comparison of the effective heat transport enhancement induced by rotation, the heat transport $\Nu(\Pr,\Ra,\iRo)$ is normalized by the one of the non-rotating case per $\Ra$ and $\Pr$: $\Nu_0(\Pr,\Ra)=\Nu(\Pr,\Ra,\iRo=0)$ \footnote{We note that due to the huge computational costs the value $\Nu_0(\Pr=6.4,\Ra=10^{10})=125.5$ is an estimate based on the other simulated $\Nu_0$ cases and the predictions by the GL theory \citep{grossmann_scaling_2000,grossmann_thermal_2001,stevens_unifying_2013}.}.\\

It is instructive to plot the effective heat transport enhancement $\Nu/\Nu_0$ in the two-dimensional (2D) parameter space of rotation $\iRo$ and thermal driving $\Ra$ to identify global trends for the optimal rotation $\iRoopt$ and the magnitude of heat transport enhancement $\Nu_{\mathrm{max}}/\Nu_0$ (Fig.~\ref{fig:NuAvg}(c,d)). Thereby, we can clearly distinguish between two behaviors, one for smaller $\Ra$ and one for larger $\Ra$ for both $\Pr$. For smaller $\Ra$, $\iRoopt$ increases with increasing $\Ra$ by following the well-known scaling for a constant boundary layer ratio \citep[e.g.,][]{julien_rapidly_1996,king_boundary_2009,yang_what_2020} (see next Sec.~\ref{subsec:BLratio} for details). For larger $\Ra$, on the contrary, $\iRoopt$ decreases again with further increasing $\Ra$. Origin and possible scaling for the large-$\Ra$ behavior of $\iRoopt$ are still unknown. In the following, we refer to the two behaviors as the \textit{small-$Ra$ subregime} and the \textit{large-$Ra$ subregime} as parts of the rotation-affected regime (in line with \citep{yang_what_2020}). The transition between the two subregimes appears to be $\Pr$-dependent at $\Ra_t\approx4\cdot10^8$ for $\Pr=4.38$ and $\Ra_t\approx7\cdot10^8$ for $\Pr=6.4$. Considering the uncertainties of the data and its sampling in the 2D parameter space, this still suggests that the transition could be given by a fixed transitional Grashof number $\Gr_t=\Ra_t/\Pr\approx10^8$. However, more data for different $\Pr$ is needed to verify this hypothesis. The second and most striking difference between the two subregimes is the amplitude of the maximal heat transport enhancement. While the maximum enhancement reaches up to $20-30\%$ in the small-$\Ra$ subregime, it decreases gradually across the transition and nearly vanishes with $\lesssim5\%$ for $\Ra=10^{10}$.

\subsection{\label{subsec:BLratio}Connection to the ratio of boundary layer thicknesses}

\begin{figure*}
\centering
\includegraphics[width=\textwidth]{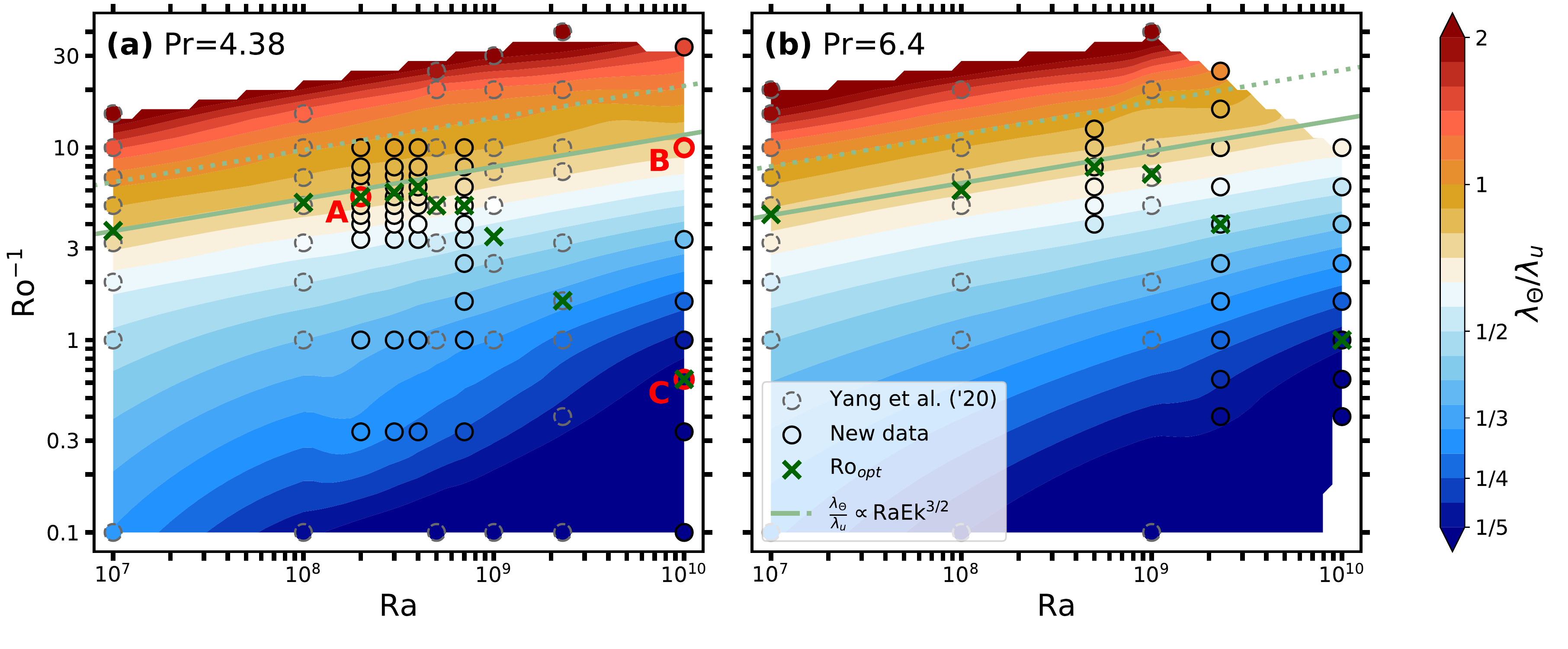}\\
\vspace{-0.3cm}
\caption{\label{fig:BLratio}Ratio of thermal and kinetic boundary layers thicknesses $\lambda_\Theta/\lambda_u$ in the $(\Ra,\iRo)$ parameter space for \textbf{(a)}~$\Pr=4.38$ and \textbf{(b)}~$\Pr=6.4$ (dashed circles - data points from \citet{yang_what_2020}, closed circles - new data points, background - cubic interpolation). The crosses mark the maximal heat transport $\Nu_{\mathrm{max}}(\Ra)$ per $\Ra$ (as in Fig.~\ref{fig:NuAvg}(c,d)). The green lines show the theoretical scaling for a constant boundary layer ratio $\lambda_\Theta/\lambda_u\approx1\propto\Ra\,\Ek^{3/2}$ with the prefactor of \citet{yang_what_2020} (solid) or \citet{king_heat_2012} (dotted). A, B, and C mark the cases presented in Fig.~\ref{fig:FlowSnaps}.}
\end{figure*}

In the common understanding, the heat transport enhancement for intermediate rotation is related to the vertical transport within vertically coherent vortices induced by Ekman pumping \citep[e.g.,][]{rossby_study_1969,julien_rapidly_1996,kunnen_heat_2006,liu_heat_2009}. Thereby, Ekman pumping is most efficient in terms of the heat transport when thermal and kinetic boundary layers have approximately equal thicknesses \citep{stevens_optimal_2010,yang_what_2020}. Assuming a kinetic Ekman-type boundary layer $\lambda_u\propto\Ek^{1/2}$ and a non-rotating thermal boundary layer that approximately scales as $\lambda_\Theta\propto\Ra^{-1/3}$ (as long $\Nu\approx\Nu_0$), one finds that a constant boundary layer ratio $\lambda_\Theta/\lambda_u$ should follow \citep{king_heat_2012}:
\begin{equation}
\lambda_\Theta/\lambda_u\propto\Ek^{3/2}\Ra
\hspace{0.2cm}\text{.}
\label{eq:BLscale}
\end{equation}
This further implies that the optimal rotation to maximize the heat transport is expected to scale as:
\begin{equation}
\iEkopt\propto\Ra^{2/3} \;\Leftrightarrow\; \iRoopt\propto\Pr^{1/2}\Ra^{1/6}
\hspace{0.2cm}\text{.}
\label{eq:optrot}
\end{equation}

To check this, we determine the thermal and kinetic boundary layer thicknesses $\lambda_{\Theta}$ and $\lambda_{u}$ by the height of the first peak in the vertical profiles of the horizontally averaged root-mean-square (RMS) temperature and horizontal velocity, respectively. This method is chosen to be in line with the previous analysis in \citet{yang_what_2020}. The reported values are averaged over top and bottom boundary layers. We find that the boundary layer ratio follows the expected scaling Eq.~\ref{eq:BLscale} throughout the entire range of $\Ra$ (Fig.~\ref{fig:BLratio}). However, only in the small-$\Ra$ subregime does the heat transport maxima correlate with this scaling such that Eq.~\ref{eq:optrot} holds. Since the general scaling properties of the boundary layer ratio do not change, the large-$\Ra$ characteristics of heat transport enhancement must originate from a different effect, which strengthens with $\Ra$ and reduces the heat transport regardless of the boundary layer ratio. The maximal heat transport is then reached at a less beneficial boundary layer ratio, just before that counteracting effect (see Secs.~\ref{sec:Flow},~\ref{sec:RaReg}) becomes dominant and truncates any further enhancement. Such a superposition behavior can also explain the decreasing trend of the enhancement magnitude.\\

\begin{figure*}
\centering
\includegraphics[width=\textwidth]{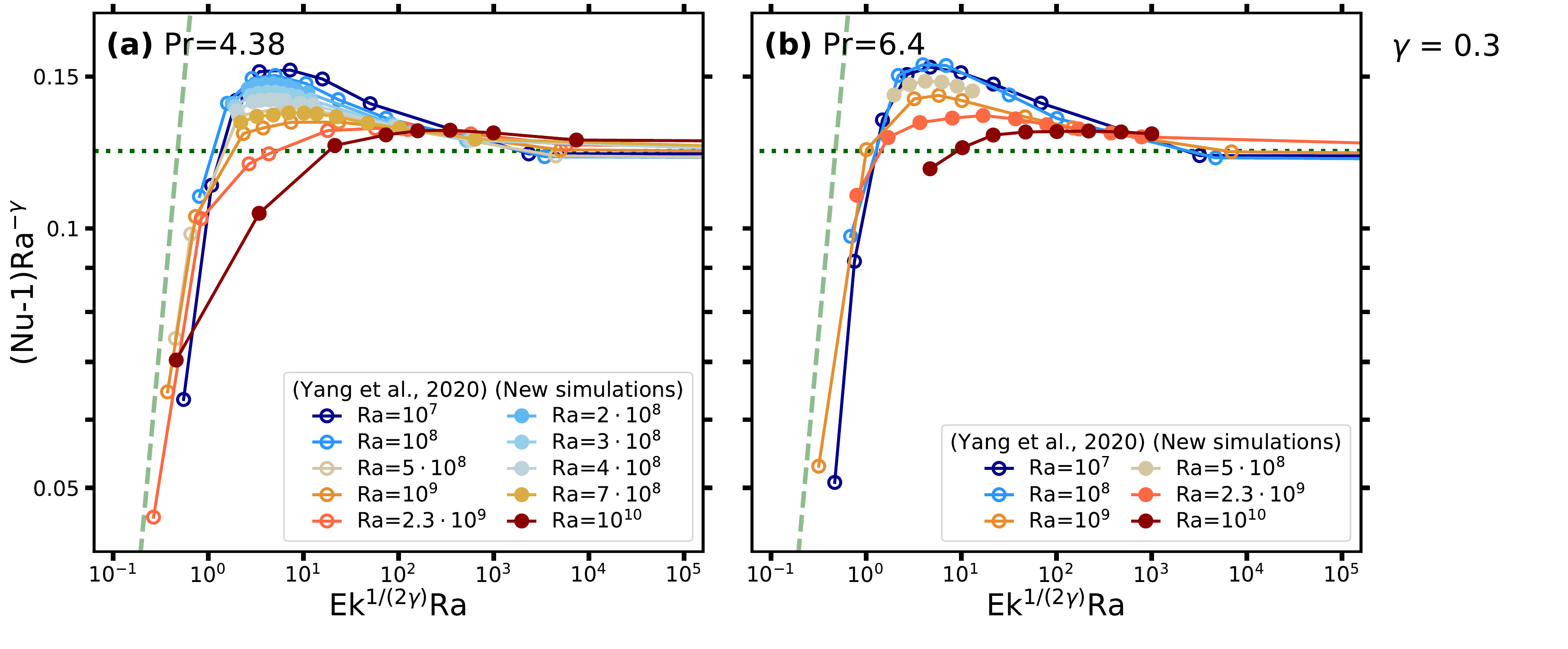}\\
\vspace{-0.3cm}
\caption{\label{fig:collapse}Collapse of the heat transport data according to the scaling theory by \citet{ecke_turbulent_2023} for $\gamma=0.3$, the exponent of the non-rotating heat transport scaling ($\Nu-1\propto\Ra^\gamma$): \textbf{(a)}~$\Pr=4.38$ and \textbf{(b)}~$\Pr=6.4$. The dotted and dashed lines indicate the expected scalings in the buoyancy-dominated regime $(\Nu-1)/\Ra^{0.3}=const$ (towards the right) and the rotation-dominated regime $(\Nu-1)/\Ra^{0.3}\propto(\Ek^{10/6}\Ra)^{s_{RD}}$ with $s_{RD}=8\gamma^2/(3-8\gamma)=1.2$ (towards the left), respectively.}
\end{figure*}

Another evidence that the global boundary layer dynamics still holds in the large-$\Ra$ subregime can be found by applying the scaling theory by \citet{ecke_turbulent_2023}. This theory provides a unifying approach for the heat transport scaling in the limits of the buoyancy- and rotation-dominated regimes. Based on the effective scaling exponent $\gamma$ in the non-rotating case ($\Nu-1\propto\Ra^\gamma$), the theory predicts the heat transport for different $\Ra$ to collapse on $(\Nu-1)\Ra^{-\gamma}$ against $\Ek^{1/(2\gamma)}\Ra$ (Fig.~\ref{fig:collapse}). Towards zero rotation, the data is supposed to show no slope. Towards strong rotation, the theory predicts a slope $s_{RD}=8\gamma^2/(3-8\gamma)$ for the data. However, the theory does not deal with the details of the transition like the heat transport enhancement. For the given $\Pr$, the effective $\gamma$ typically increases with increasing $\Ra$ approaching $\gamma=1/3$, until the expected onset of the ultimate regime for very large $\Ra\gg10^{14}$ \citep{grossmann_scaling_2000,grossmann_thermal_2001,stevens_unifying_2013}. By fitting our $\Nu_0$ data, we obtain a $\gamma\approx0.3$ across the entire range of $\Ra$ ($\gamma_{\Pr=4.38}=0.310\pm0.003$, $\gamma_{\Pr=6.4}=0.302\pm0.003$). Indeed, all $\Nu$ data show a pretty decent collapse towards the buoyancy- and rotation-dominated regimes for $\gamma=0.3$ (Fig.~\ref{fig:collapse}). We note that our data do not reach far into the rotation-dominated regime, and thus the expected slope $s_{RD}=8\gamma^2/(3-8\gamma)=1.2$ is not fully obtained in all cases. Further, one can observe that for small $\Ra$ the heat transport maxima collapse onto a fixed value $\Ek^{1/(2\gamma)}\Ra=\Ek^{10/6}\Ra=const$. This closely resembles the predicted scaling of a constant boundary layer ratio $\Ek^{3/2}\Ra=const$ (Eq.~\ref{eq:BLscale}), which is obtained by assuming $\gamma=1/3$. On the contrary, the data of large $\Ra$ only collapse in the limits of the buoyancy- and rotation-dominated regimes. But still their heat transport maxima neither collapse nor simply line up along a fixed $\Ek^{1/(2\gamma)}\Ra$. The heat transport enhancement reaches an upper bound for small $\Ra$, which depicts the potential enhancement due to the influence of the boundary layer ratio. With increasing $\Ra$, the enhancement peak gets ``peeled'' such that the heat transport maximum retreats to slower rotation and looses efficiency.\\

One now could suppose that accounting for the changing exponent $\gamma$ could help to collapse all the data. Therefore, we also checked the robustness of the above collapse. Considering only the data in the small $\Ra$ and large-$\Ra$ subregimes, we obtain $\gamma=0.294\pm0.001$ and $\gamma=0.316\pm0.003$, respectively. These minor differences can improve the collapse of the small-$\Ra$ or large-$\Ra$ data in the limits of the buoyancy- and rotation-dominated regimes, respectively, but do not resolve the offset of the heat transport maxima at large $\Ra$ (see Supplemental Material \citep{suppmat}, Fig.~S2). The same applies for the theoretical value of $\gamma=1/3$. This, however, is not surprising since the above scaling theory \citep{ecke_turbulent_2023} essentially builds on fixed scaling relations for the boundary layers and their ratio over the entire range of $\Ra$, while the heat transport maxima at large-$\Ra$ start to deviate from it (Fig.~\ref{fig:BLratio}).\\

In brief, we identified that fundamental scaling properties for the thickness ratio of the thermal and kinetic boundary layers hold across the entire range of $\Ra$. The boundary layer ratio controls the upper bound of the heat transport enhancement as well as the collapse of the data towards the limits of the buoyancy- and rotation-dominated regimes. Albeit, the heat transport must be additionally affected by another effect, which strengthens with $\Ra$ suppressing enhancement. The remaining sections aim to further reveal this additional effect.

\section{\label{sec:Flow}Flow morphology at small and large Rayleigh numbers}

\begin{figure*}
\centering
\includegraphics[width=\textwidth]{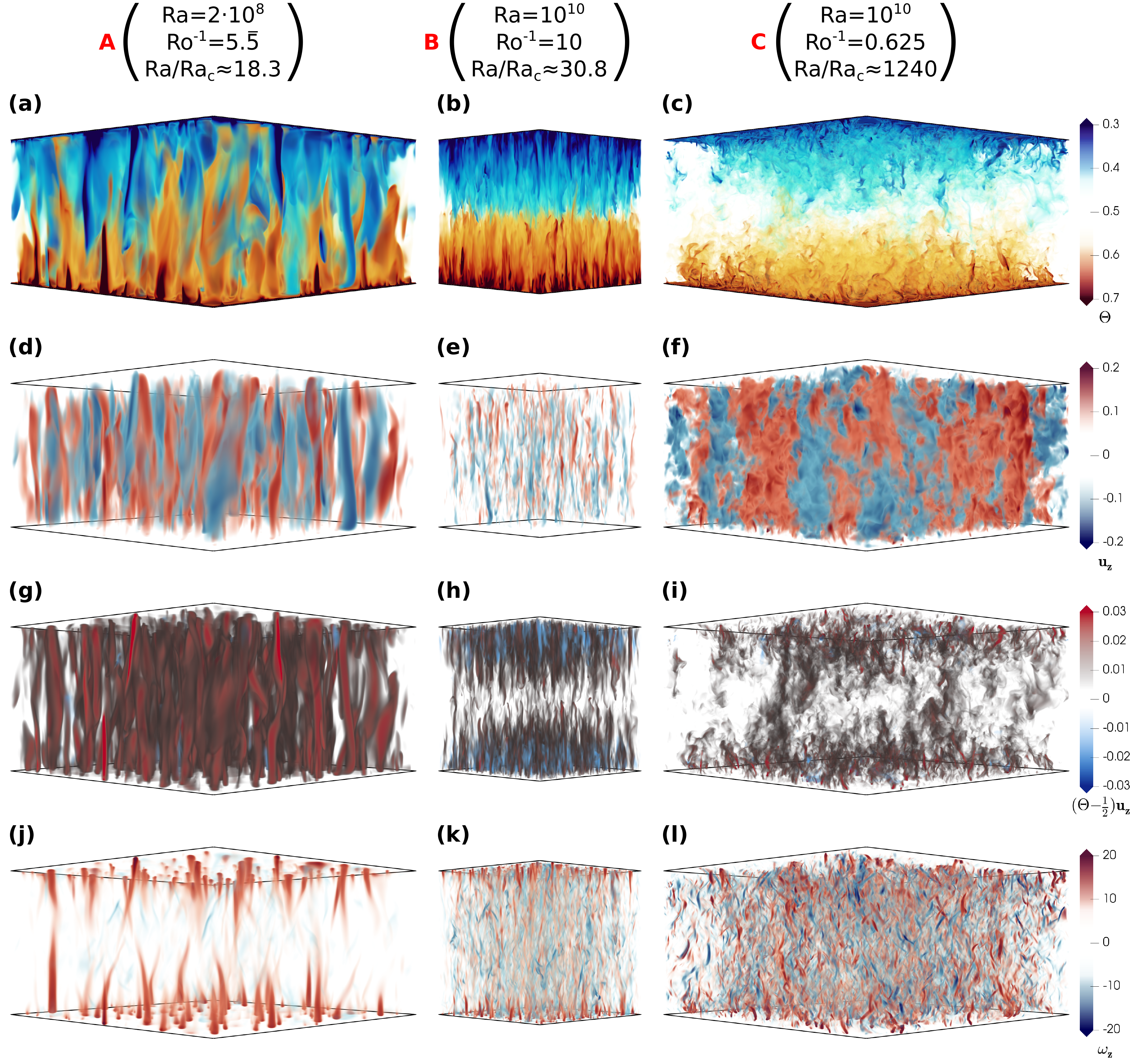}
\caption{Instantaneous snapshots of characteristic flow quantities for case A, the heat transport maximum for $\Ra=2\cdot10^8$ ($\iRo=5.\bar{5}$, left column), for case B, the expected heat transport maximum for $\Ra=10^{10}$ ($\iRo=10$, middle column), and for case C, the observed heat transport maximum for $\Ra=10^{10}$ ($\iRo=0.625$, right column), respectively, all at $\Pr=4.38$: \textbf{(a-c)}~the temperature field~$\Theta$, \textbf{(d-f)}~the vertical velocity~$u_z$, \textbf{(g-i)}~the local convective heat transport~$(\Theta-\frac{1}{2})\,u_z$, and \textbf{(j-l)}~the vertical vorticity~$\omega_z=\partial_x u_y - \partial_y u_x$.}
\label{fig:FlowSnaps}
\end{figure*}

Next, we take a closer look at the flow morphology that yields the maximal heat transport at small and large $\Ra$. We will consider three cases of $\Pr=4.38$: (A) in the small-$\Ra$ subregime ($\Ra=2\cdot10^8$), where the enhancement maximum is observed at the expected rotation rate ($\iRo=5.\bar{5}$) when $\lambda_\Theta/\lambda_u\approx1$, (B) in the large-$\Ra$ subregime ($\Ra=10^{10}$), where no enhancement maximum is observed but would be expected based on $\lambda_\Theta/\lambda_u\approx1$ ($\iRo=10$), and (C) where the enhancement maximum is actually observed ($\iRo=0.625$) for the very same $\Ra=10^{10}$ in the large-$\Ra$ subregime. Comparing the instantaneous temperature, vertical velocity, local convective heat flux, and vertical vorticity fields reveals the striking differences between the flow in the small-$\Ra$ and large-$\Ra$ subregimes (see Fig.~\ref{fig:FlowSnaps}).\\

For case A in the small-$\Ra$ subregime, the flow is predominantly organized in domain-spanning, vertically coherent vortices (Fig.~\ref{fig:FlowSnaps}(j)) with Ekman pumping in their interior (Fig.~\ref{fig:FlowSnaps}(d)), which effectively transports hot and cold fluid through the bulk (Fig.~\ref{fig:FlowSnaps}(a)) and leads to a path of very high vertical local heat flux shortcutting the bulk region (Fig.~\ref{fig:FlowSnaps}(g)). Such a flow organization can benefit massively from an ideal boundary layer ratio, and thus, the observed heat transport maximum corresponds to its expected ($\Ra,\iRo$) location based on the scaling of the boundary layer ratio.\\

The flow morphology is completely different at large $\Ra$. For case B, where the maximum is expected based on $\lambda_\Theta/\lambda_u\approx1$, the vertically coherent vortices get disrupted in the bulk (Fig.~\ref{fig:FlowSnaps}(k)) as the vortices of smaller width are easier perturbed by the stronger turbulence. This results in less-coherent structures in the vertical velocity (Fig.~\ref{fig:FlowSnaps}(e)), which interrupt the Ekman pumping from the boundary layers. Further, these remaining vortex structures show a leakage of hot (cold) fluid in the lower (upper) half of the domain (Fig.~\ref{fig:FlowSnaps}(b)), which leads to a strong non-zero temperature gradient in the bulk. The vertical heat transport is less focused within these structures compared to A, and the leaking fluid can locally create significant negative contributions to the overall heat transport (Fig.~\ref{fig:FlowSnaps}(h)). Altogether, the bulk flow has changed in a way that Ekman pumping becomes ineffective and cannot benefit anymore from an ideal boundary layer ratio.\\

We now come to case C, i.e., the heat transport maximum at large $\Ra$, which occurs at relatively weak rotation rates. Therefore, the bulk looks still rather buoyancy-dominated with $\iRoopt=0.625$. The bulk flow is characterized by a fully decorrelated small-scale vorticity (Fig.~\ref{fig:FlowSnaps}(l)) and much larger turbulent structures in the vertical velocity (Fig.~\ref{fig:FlowSnaps}(f)), which differ from classical Ekman pumping within elongated Ekman vortices. Nonetheless, one can identify spiraling plumes of hot or cold fluid, i.e., plumes that already show some vortical component, next to the plates (Fig.~\ref{fig:FlowSnaps}(c) vs. (l)). Similarly to case A, they show a very high vertical local heat flux (Fig.~\ref{fig:FlowSnaps}(i)). However, they do not reach far into the bulk, and thus their contribution to the overall heat transport is less relevant. The heat transport through the bulk is rather related to the large-scale velocity structure than to the small-scale vorticity structures.\\

A better understanding of the flow morphology and the underlying scales can be gained from the radial spectra of the flow quantities in Fig.~\ref{fig:FlowSnaps}. The time-averaged horizontal spectra $\Phi_{q_1,q_2}^{2D}$ of the quantity $q_{1,2}\in\{u_z,\Theta,\omega_z\}$ at a given height $z$ is given by:
\begin{equation}
\begin{split}
\Phi_{q_1,q_2}^{2D}(k_x,k_y)&=\left\langle\mathbb{R}\mathrm{e}\left( \mathcal{F}[q_1(x,y)]\cdot\mathcal{F}[q_2(x,y)]^\ast\right)\right\rangle_{N_t}
\hspace{0.2cm}\text{,}\\
\mathcal{F}(q(x,y))&=\sum\limits_{k_x,k_y} \hat{q}(k_x,k_y) \cdot \exp(ik_x x +ik_y y)
\hspace{0.2cm}\text{.}
\end{split}
\end{equation}
Here, $k_x$ and $k_y$ denote the discrete wavenumbers, $\mathcal{F}$ describes a discrete Fourier transformation in both horizontal directions decomposing $q$ into its coefficients $\hat{q}$, while $\mathbb{R}\mathrm{e}$ and $^\ast$ refer to the real part and the complex conjugate of a complex number, respectively. Temporal averaging is achieved by collecting $N_t$ snapshots of the 3D flow fields after the flow reached its statistically stationary state. Additionally, the spectra make use of the top-bottom symmetry and are averaged over the top and bottom half at similar distance from midheight. For computing the radial spectra $\Phi(k_r)$, the 2D spectra $\Phi^{2D}$ is first interpolated to a polar $(k_r,\vartheta)$ grid in order to improve the integration along the discrete values of $k_r=\sqrt{k_x^2+k_y^2}$. We then calculate the radial spectra $\Phi_{q_1,q_2}$ as follows:
\begin{equation}
\Phi_{q_1,q_2}(k_r)=\int k_r\;\Phi_{q_1,q_2}^{2D}(k_r,\vartheta)\;\mathrm{d}\vartheta
\hspace{0.2cm}\text{.}
\end{equation}
In this way, the (co)variance of $\left<q_1 q_2\right>_\mathcal{H}$ is conserved by linear integration over $k_r$:
\begin{equation}
\left<q_1 q_2\right>_\mathcal{H}=\iint \Phi_{q_1,q_2}^{2D}(k_x,k_y)\;\mathrm{d}k_x\mathrm{d}k_y=\int\Phi_{q_1,q_2}(k_r)\;\mathrm{d}k_r
\hspace{0.2cm}\text{.}
\end{equation}

\begin{figure*}
\centering
\includegraphics[width=\textwidth]{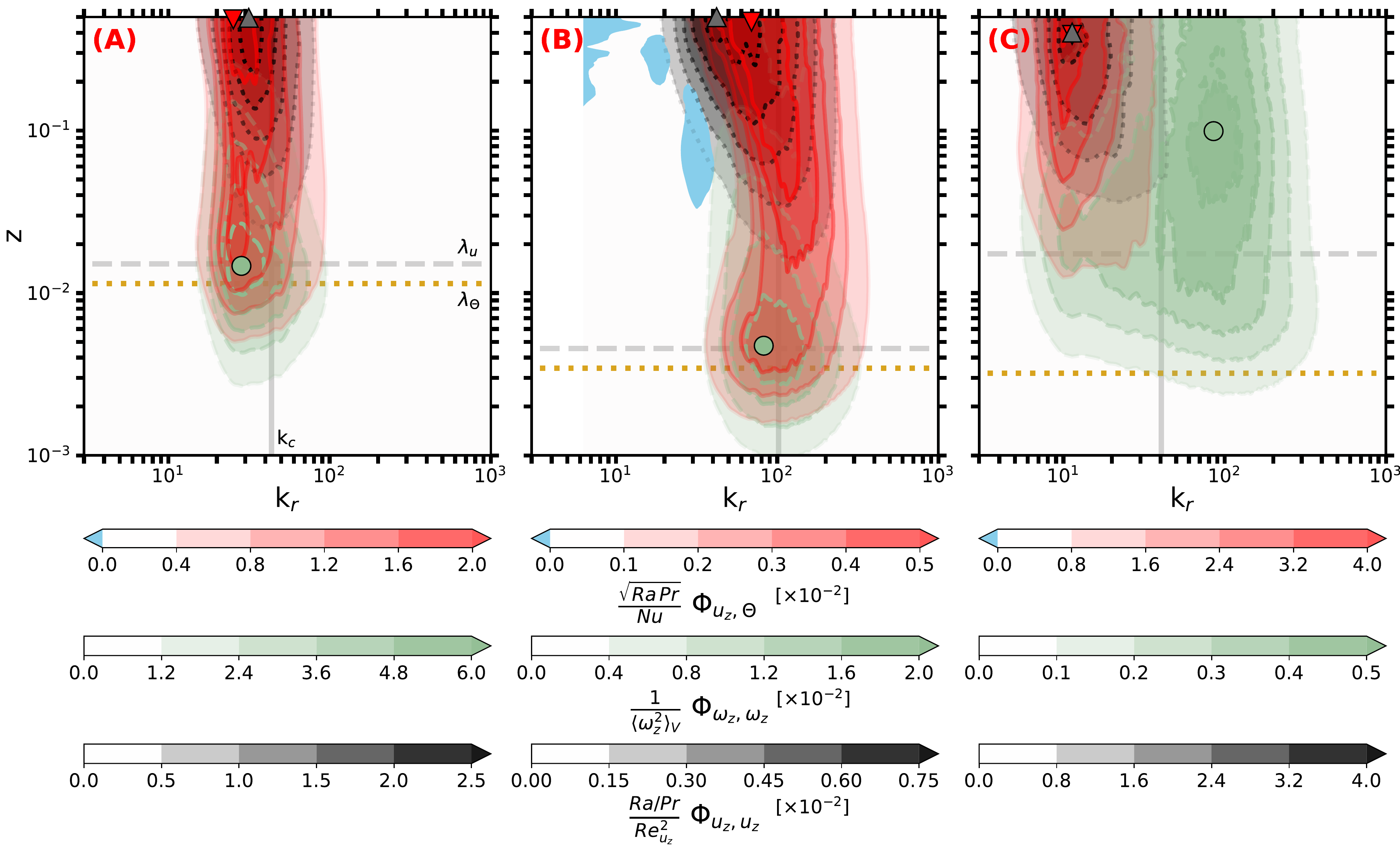}
\caption{\label{fig:kspectra}Time-averaged, radial spectra $\Phi_{q_1,q_2}$ of the vertical velocity ($q_{1,2}=u_z$), the vertical vorticity ($q_{1,2}=\omega_z$), and the local heat transport (cospectra $q_1=u_z, q_2=\Theta$) as function of $z$ and $k_r$ for the example cases A, B, and C, see Fig.~\ref{fig:FlowSnaps}. Upward triangle, dot, and downward triangle mark the peak location of $\Phi_{u_z,u_z}$, $\Phi_{\omega_z,\omega_z}$, and $\Phi_{u_z,\Theta}$, respectively. The dashed and dotted lines are kinetic and thermal boundary layer thicknesses. The vertical solid line scales with $k_c=\frac{2\pi}{4.82}\Ek^{-1/3}$ and serves as a guide to the eye for the theoretical size of horizontal structures in rotating RBC.}
\end{figure*}

In these radial spectra $\Phi_{q_1,q_2}$ (Fig.~\ref{fig:kspectra}; the corresponding premultiplied spectra $k_r\Phi_{q_1,q_2}$ are shown in the Supplemental Material \citep{suppmat}, Fig.~S3), we find that for case A not only the boundary layer ratio is beneficial, but also the predominant horizontal scales of vertical vorticity $\omega_z$, vertical velocity $u_z$, and local convective heat transport $(\Theta-0.5)u_z$ correlate throughout the entire domain. The peak of $\Phi_{\omega_z,\omega_z}$ is located at the height of the kinetic boundary layer, depicting the predominant spacing of the emerging Ekman vortices. The peak of $\Phi_{u_z,u_z}$ relates to the vertically coherent transport induced by Ekman pumping. Its amplitude increases within the bulk and peaks at midheight. As expected, the heat transport cospectrum $\Phi_{u_z,\Theta}$ connects the prior two at roughly one wavelength through the entire domain. This yields an efficient heat transport through the bulk, resulting in the large enhancement at small $\Ra$, like case A.\\

The favorable match of the different horizontal scales is lost when the boundary layer ratio becomes most effective for $\Ra=10^{10}$ (Fig.~\ref{fig:kspectra}(B)). The peak of $\Phi_{\omega_z,\omega_z}$ shifts to larger $k_r$, meaning that the spacing of Ekman vorticies becomes denser as the Ekman number decreases ($k_c=2\pi/l_c\propto\Ek^{-1/3}$). Concurrently, the peak of $\Phi_{u_z,u_z}$ remains at larger wavelength and broadens creating a significant offset $k_{u_z}^\mathrm{max}<k_{\omega_z}^\mathrm{max}$. The heat transport $\Phi_{u_z,\Theta}$ just above the kinetic boundary layer still correlates with the Ekman vortices $\Phi_{\omega_z,\omega_z}$. Further into the bulk, $\Phi_{u_z,\Theta}$ deflects, first towards smaller wavelengths. At midheight, the peak of $\Phi_{u_z,\Theta}$ remains at smaller scales than the predominant vertical velocity structures $\Phi_{u_z,u_z}$. Furthermore, we observe significant portions with negative contributions to the heat transport in the central bulk at relatively large wavelengths. We conjecture that these three factors, (i) mismatch of vortex and vertical velocity structures, (ii) heat transport along varying wavelengths, and (iii) negative heat transport at larger wavelengths, are consequences of the denser spaced Ekman vortices being less robust against perturbations from increased turbulent fluctuations. Each of the three reduces the efficiency of heat transport through the bulk, which leads to a reduction of the overall heat transport, despite of the beneficial boundary layer ratio.\\

For case C, the observed heat transport maximum for $\Ra=10^{10}$, the vorticity spectra $\Phi_{\omega_z,\omega_z}$ appears rather broad at small wavelength with its primary peak above the boundary layers (Fig.~\ref{fig:kspectra}(C)). This peak is related to the small-scale turbulent fluctuations in the buoyancy-dominated bulk. Besides the major peak, a secondary peak starts to form in $\Phi_{\omega_z,\omega_z}$ at the kinetic boundary layer height. We relate this to the onset of Ekman vortex formation in rotation-affected regime, although located at a larger scale than suggested by the $l_c\propto\Ek^{1/3}$ scaling, which presumably does not hold towards the buoyancy-dominated regime. Interestingly, the secondary vorticity peak occurs at the same wavelength as the predominant velocity structures $\Phi_{u_z,u_z}$, which again provides a preferable wavelength for the heat transport through the bulk. Indeed, $\Phi_{u_z,\Theta}$ peaks and elongates along this wavelength. The agreement of the horizontal scales is most likely the reason why still some heat transport enhancement can be observed, although the boundary layer ratio is far away from being optimal.\\

\begin{figure*}
\centering
\includegraphics[width=\textwidth]{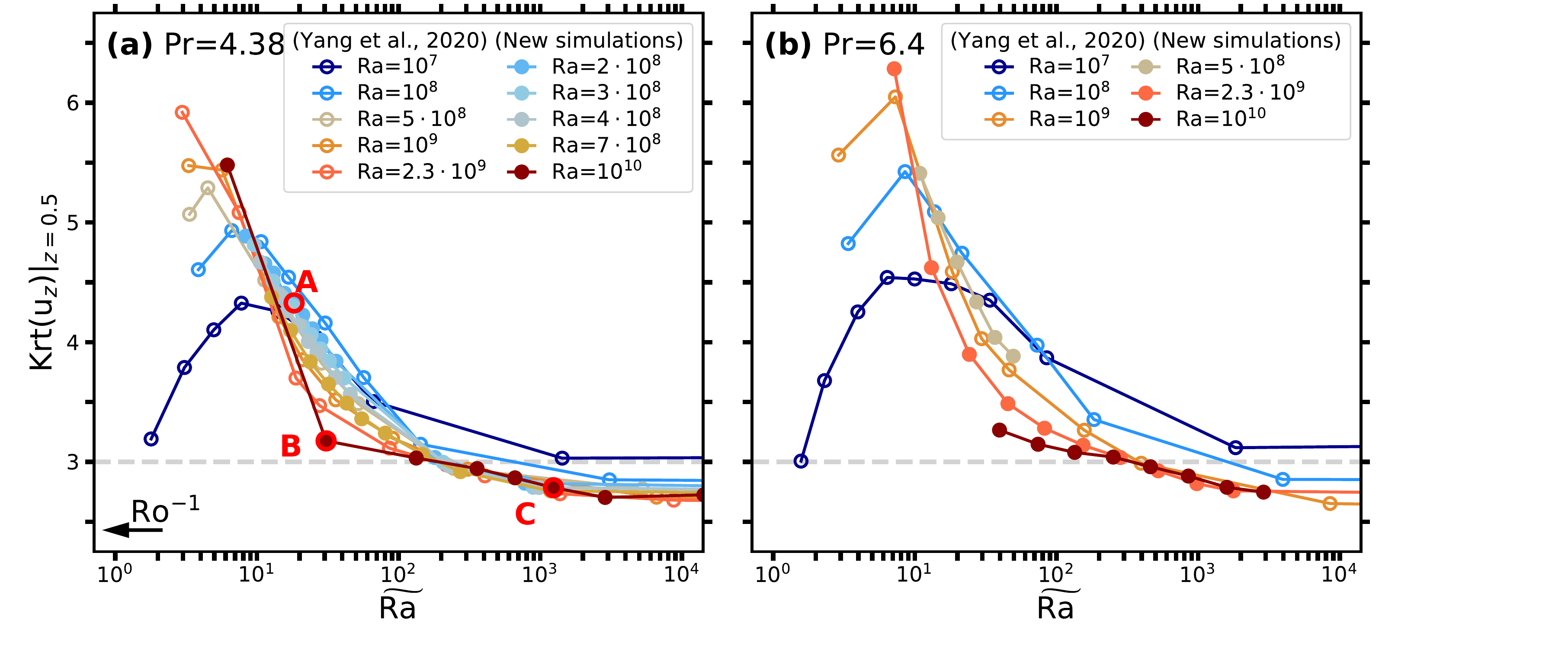}\\
\includegraphics[width=\textwidth]{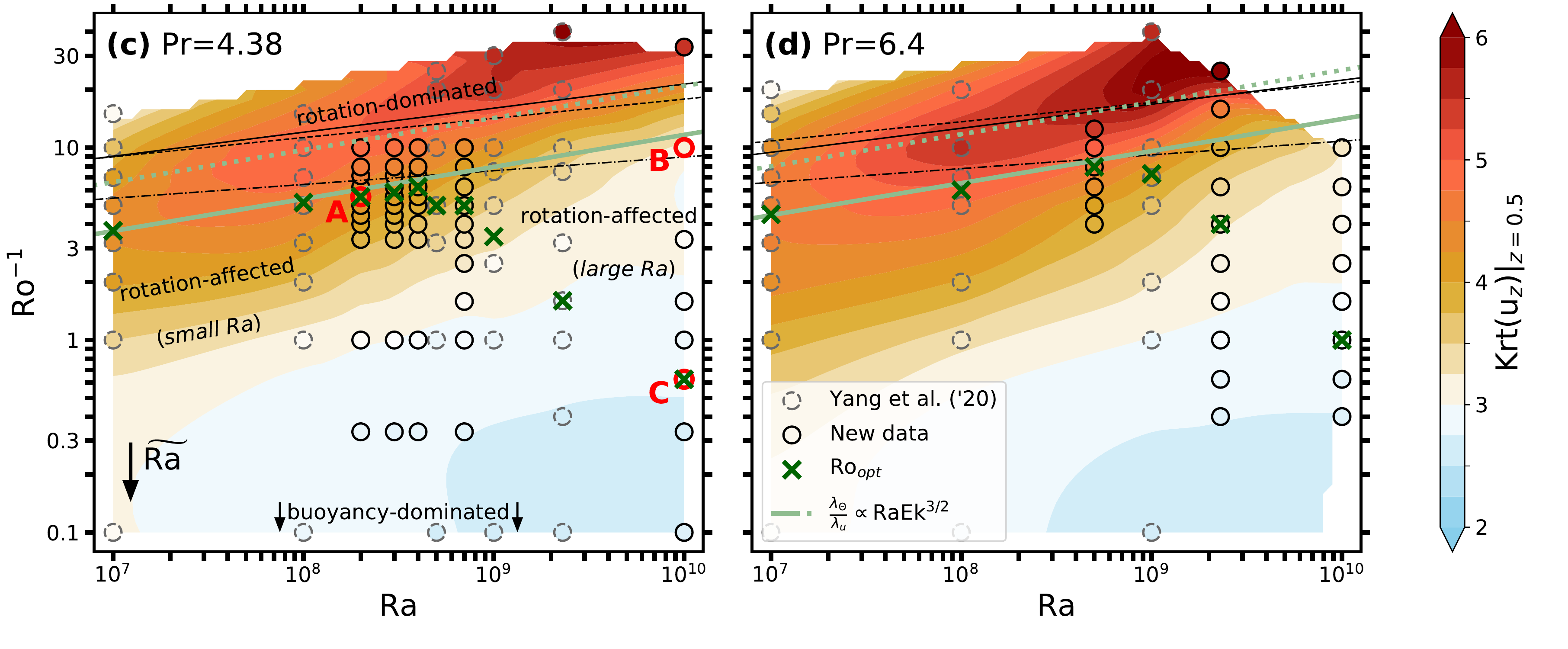}\\
\vspace{-0.3cm}
\caption{\label{fig:KrtUZ}Kurtosis of the vertical velocity $u_z$ at midheight ($z=0.5$): \textbf{(a,b)}~as the function of $\widetilde{\Ra}=\Ra/\Ra_c$ and \textbf{(c,d)}~in the $(\Ra,\iRo)$ parameter space for $\Pr=4.38$ and $\Pr=6.4$, respectively (dashed circles - data points from \citet{yang_what_2020}, closed circles - new data points, background - cubic interpolation). The crosses mark the maximal heat transport $\Nu_{\mathrm{max}}(\Ra)$ per $\Ra$ (as in Fig.~\ref{fig:NuAvg}(c,d)). The green lines scale as $\lambda_\Theta/\lambda_u\approx1\propto\Ra\,\Ek^{3/2}$ with the prefactor of \citet{yang_what_2020} (solid) or \citet{king_heat_2012} (dotted), indicating the transition between rotation-affected and rotation-dominated regimes. The thin black lines show alternative definitions for this transition by \citet{julien_heat_2012} (solid), \citet{ecke_heat_2014} (dashed), or \citet{king_boundary_2009} (dashed-dotted). A, B, and C mark the cases presented in Fig.~\ref{fig:FlowSnaps}. In (c), the main flow regimes are indicated as orientation for the reader.}
\end{figure*}

From these spectra, we conclude that the mismatch of the horizontal scales is the major reason for the cutoff of the heat transport enhancement in the large-$\Ra$ subregime. A mismatch between the spacing of the Ekman vortices above boundary layer height and the structures for vertical transport in the bulk reduces the efficiency with which heat can be transported through the bulk. The mismatch reflects a loss of coherence due to the observed transition in the flow morphology from domain-spanning vertically coherent vortices at small $\Ra$ to decorrelated vortical structures in the bulk at large $\Ra$. These short decorrelated vortical structures have a remarkable similarity to the flow state of geostrophic turbulence in the rotation-dominated regime \citep{julien_statistical_2012,nieves_statistical_2014,stellmach_approaching_2014}. A common feature of this flow state is the formation of barotropic large-scale vortex condensates in the bulk \citep{stellmach_approaching_2014,rubio_upscale_2014,aguirre-guzman_competition_2020,aguirre-guzman_force_2021}. However, they typically form under free-slip boundaries and are rarely observed for no-slip boundaries \citep{stellmach_approaching_2014,kunnen_transition_2016,plumley_effects_2016}. Therefore, we do not observe such a large-scale vortex in cases B or C at $\Ra=10^{10}$, in agreement with \citet{kunnen_transition_2016} for similarly large $\Ra$.\\

The geostrophic turbulence flow state is typically observed at relatively large values of $\widetilde{\Ra}=\Ra/\Ra_c$ (e.g., $\widetilde{\Ra}\gtrsim18$ for $\Pr=3$ \citep{julien_statistical_2012}), whereas at smaller $\widetilde{\Ra}$ the flow remains in the vortical \textit{plume state} with long vertically coherent vortices \citep[][Fig.~2(a) therein]{julien_statistical_2012}. Only for sufficiently large $\Ra$ does $\widetilde{\Ra}$ become large enough within the rotation-dominated regime to trigger the geostrophic turbulence flow state. For smaller $\Ra$, the transition into the rotation-affected regime remains out off the vortical \textit{plume state}. We note that previously the full range of these flow states have been observed experimentally \citep{cheng_laboratory-numerical_2015,cheng_heuristic_2018}; however, numerically they have been studied only deep in the rotation-dominated regime, i.e., for $\Ek<10^{-6}$ mostly by the reduced set of equations for rapidly rotating RBC \citep{sprague_numerical_2006,julien_statistical_2012}. Although the reduced equations loose their validity outside the rotation-dominated regime, we think that similar subsequent flow states must still persist in the rotation-affected regime, depending on the value of $\widetilde{\Ra}$ that is reached at that transition.\\

According to \citet{julien_statistical_2012}, the flow reaches the geostrophic turbulence state, when the kurtosis of the vertical velocity at midheight reapproaches the Gaussian value of $\mathrm{Krt}(u_z)=3$ with increasing $\widetilde{\Ra}$ \citep[][Fig.~2(a,b) therein]{julien_statistical_2012}. Computing the midheight kurtosis for our cases, we observe the same trend (Fig.~\ref{fig:KrtUZ}). Outside the rotation-dominated regime, i.e., large $\widetilde{\Ra}\to\infty$, the kurtosis decreases further to $\mathrm{Krt}(u_z)\approx2.7<3$ \citep{aguirre-guzman_competition_2020,aguirre-guzman_flow-_2022}. Thus, for our cases, $\mathrm{Krt}(u_z)\approx3$ can generally indicate the following: (i) the presence of geostrophic turbulence if inside or close to the rotation-dominated regime, or (ii) a transition to non-rotating convection if inside or close to the buoyancy-dominated regime. In the small-$\Ra$ subregime (e.g., case A), $\mathrm{Krt}(u_z)|_{z=0.5}>3$ indicates vertically coherent vortices when $\lambda_\Theta/\lambda_u$ is most beneficial. In the large-$\Ra$ subregime, the kurtosis decreases along the optimal $\lambda_\Theta/\lambda_u$ down to $\mathrm{Krt}(u_z)\approx3$ at $\Ra=10^{10}$ (case B). Accordingly, the flow at the transition to the rotation-affected regime has (almost) reached the geostrophic turbulence state, which impedes efficient vertical transport of heat through the bulk due to the breakup of the coherent vortices fed by Ekman pumping at boundary layer height. This also implies that the flow undergoes a different transition in the rotation-affected regime for small $\Ra$ (from plumelike vortices to large-scale circulation) than for large $\Ra$ (from geostrophic turbulence to large-scale circulation).\\

Another characteristic of the geostrophic turbulence flow state is a relatively strong non-zero vertical temperature gradient in the bulk \citep{julien_statistical_2012,stellmach_approaching_2014}. In the rotation-dominated regime, the bulk temperature gradient $\partial_z\Theta\vert_\text{bulk}$ first increases with increasing $\widetilde{\Ra}$ from $-1$ at the onset of convection towards $0$. However, depending on $\Pr$, it decreases again and saturates at a significant non-zero gradient ($\approx-0.4$ in \citep{julien_statistical_2012}) when the flow reaches the geostrophic turbulence state. \citet{julien_statistical_2012} did not address how the gradient must further increase to zero when the flow transitions from geostrophic turbulence at large $\widetilde{\Ra}$ to buoyancy-dominated convection with $\partial_z\Theta\vert_\text{bulk}\approx0$ at even larger $\widetilde{\Ra}\to\infty$. First evidence for the trend of bulk gradient from geostrophic turbulence towards buoyancy-dominated convection are presented in \citet{aguirre-guzman_flow-_2022}.\\

\begin{figure*}
\centering
\includegraphics[width=\textwidth]{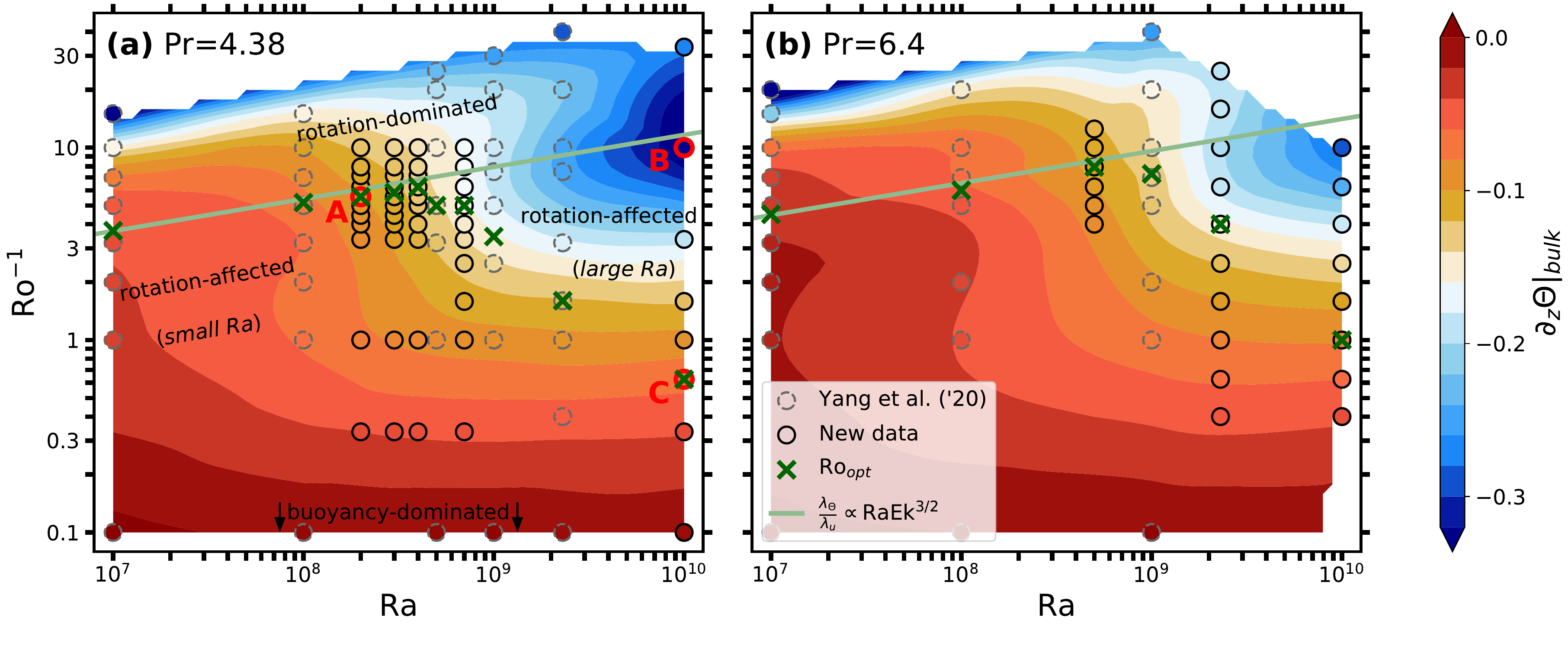}\\
\vspace{-0.3cm}
\caption{\label{fig:Tgrad}Mean vertical temperature gradient in the bulk $\partial_z\Theta\vert_\text{bulk}$ in the $(\Ra,\iRo)$ parameter space for \textbf{(a)}~$\Pr=4.38$ and \textbf{(b)}~$\Pr=6.4$ (dashed circles - data points from \citet{yang_what_2020}, closed circles - new data points, background - cubic interpolation). The crosses mark the maximal heat transport $\Nu_{\mathrm{max}}(\Ra)$ per $\Ra$ (as in Fig.~\ref{fig:NuAvg}(c,d)). The solid line scales as $\lambda_\Theta/\lambda_u\approx1\propto\Ra\,\Ek^{3/2}$ \citep{king_heat_2012,yang_what_2020}, indicating the transition between rotation-affected and rotation-dominated regimes. A, B, and C mark the cases presented in Fig.~\ref{fig:FlowSnaps}. In (a), the main flow regimes are indicated as orientation for the reader.}
\end{figure*}

We compute the vertical temperature gradient in the bulk as the mean vertical gradient of the time- and horizontally averaged temperature $\Theta$ between $0.2\leq z\leq0.8$: $\partial_z\Theta\vert_\text{bulk}\equiv\langle\partial_z\langle\Theta\rangle_\mathcal{H}\rangle_{z\in[0.2,0.8]}$. At small $\Ra$, where the flow directly transforms from buoyancy-dominated convection to the plume-vortex state with increasing $\iRo$, the bulk gradient remains almost zero $\partial_z\Theta\vert_\text{bulk}\gtrsim-0.1$ throughout the rotation-affected regime and continuously decreases in the rotation-dominated regime beyond the maximal heat transport (Fig.~\ref{fig:Tgrad}). Thereby, the flat bulk gradient at the heat transport maxima results from the mutually strong temperature anomalies in the vertically coherent vortices balancing in the horizontal average rather than from a homogeneously mixed bulk as in the buoyancy-dominated cases. On the contrary, we observe that the bulk gradient decreases significantly across the rotation-affected regime ($\partial_z\Theta\vert_\text{bulk}\lesssim-0.2$) for large $\Ra$. Especially for our largest $\Ra$ values $10^{10}$ and $2.3\cdot10^9$, we then observe a small increase of $\partial_z\Theta\vert_\text{bulk}$ before decreasing further in the rotation-dominated regime (Fig.~\ref{fig:Tgrad}), which agrees with the trend in \citep[][Fig.~2(a) therein]{aguirre-guzman_flow-_2022}. The resulting local minima (per $\Ra$) are another evidence for reaching the geostrophic turbulence state towards the rotation-dominated to rotation-affected transition. The change in the behavior of $\partial_z\Theta\vert_\text{bulk}$ agrees with the breakdown of heat transport enhancement and the deviating maxima locations. Therefore, we can refine our previous definitions of the two rotation-affected subregimes: the small-$\Ra$ subregime characterized by vertically coherent vortices and the large-$\Ra$ subregime characterized by small-scale \textit{rotation-affected turbulence} (as complement to geostrophic turbulence in the rotation-dominated regime).\\

Finally, we want to emphasize that the capability to enhance the heat transport is always referring to the non-rotating case $\Nu_0$ per $\Ra$. Considering those absolute values $\Nu_0$, one recognizes that, naturally, the more turbulent, mixed bulk with decorrelated small-scale vorticity at larger $\Ra$ already provides a more efficient way to transport the heat through the bulk than at smaller $\Ra$: $\Nu_0(\Pr=4.38,\Ra=10^{10})=126.38$ vs. $\Nu_0(\Pr=4.38,\Ra=2\cdot10^{8})=37.85$. In this sense, a lower base level $\Nu_0$ is easier to enhance than a higher one \citep{chong_confined_2017}.

\section{\label{sec:RaReg}Maximal heat transport enhancement at large Rayleigh numbers}

\begin{figure*}
\centering
\includegraphics[width=\textwidth]{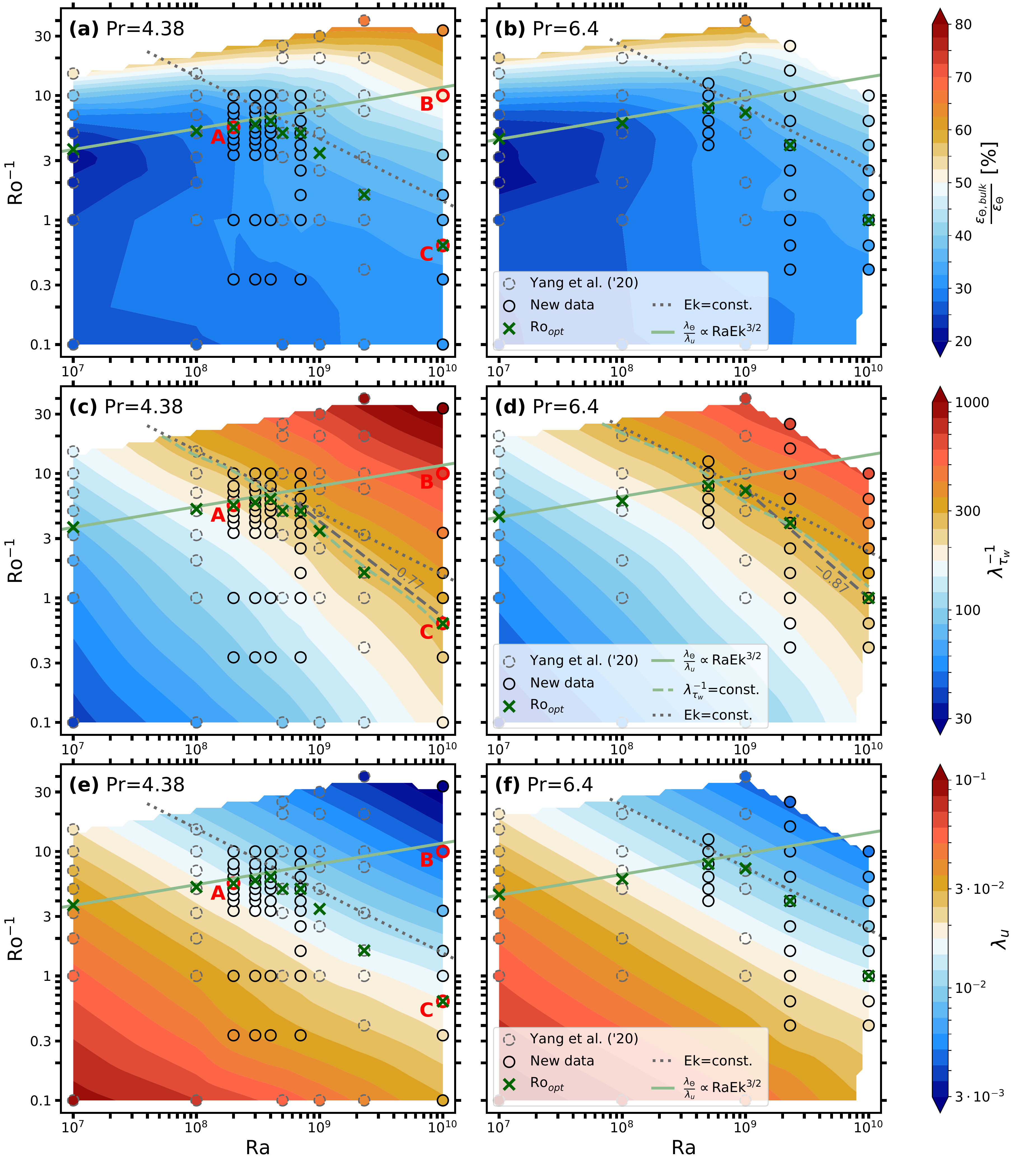}\\
\vspace{-0.3cm}
\caption{\label{fig:BLkin}\textbf{(a,b)}~Fraction of thermal dissipation in the bulk $\frac{\epsilon_{\Theta,\text{bulk}}}{\epsilon_{\Theta}}$, \textbf{(c,d)}~ratio of wall shear stress to vortex strength in terms of the inverse \textit{slope} definition for the kinetic boundary layer $\lambda_{\tau_w}^{-1}=\tau_w/U_V$ (Eq.~\ref{eq:BLslope}), and \textbf{(e,f)}~kinetic boundary layer thickness~$\lambda_{u}$ based on the classical RMS definition in the $(\Ra,\iRo)$ parameter space for $\Pr=4.38$ and $\Pr=6.4$, respectively (dashed circles - data points from \citet{yang_what_2020}, closed circles - new data points, background - (a,b) linear, (c-f) cubic interpolation). The crosses mark the maximal heat transport $\Nu_{\mathrm{max}}(\Ra)$ per $\Ra$ (as in Fig.~\ref{fig:NuAvg}(c,d)). The solid line scales as $\lambda_\Theta/\lambda_u\approx1\propto\Ra\,\Ek^{3/2}$ \citep{king_heat_2012,yang_what_2020}. The dashed lines in (c,d) emphasize contours of constant $\lambda_{\tau_w}^{-1}$ (green) and their fitted slopes (gray). The dotted lines indicate constant $\Ek$ for comparison. A, B, and C mark the cases presented in Fig.~\ref{fig:FlowSnaps}.}
\end{figure*}

In this section we look for quantities that correlate with the heat transport maxima at large $\Ra$ and therefore might control or at least indicate the optimal rotation rate $\iRoopt$ (or $\iEkopt$) in the large-$\Ra$ subregime. In Sec.~\ref{sec:HTEnBL} we hypothesized that a second effect strengthens with increasing $\Ra$ and $\iRo$, which drastically reduces the heat transport cutting off any further enhancement due to a more beneficial boundary layer ratio beyond a certain threshold. Although the steeper temperature gradient in the bulk is a clear sign for the heat leakage of the vortices and the reduction of the heat transport in the large-$\Ra$ subregime (as discussed in Sec.~\ref{sec:Flow}), it is not simply a threshold of $\partial_z\Theta\vert_\text{bulk}$ that determines the cutoff (Fig.~\ref{fig:Tgrad}).\\

The ability of the Ekman vortices to maintain the heat in their interior is rather related to the thermal dissipation in the bulk. We compute the thermal dissipation from the temperature gradients $\epsilon_\Theta=\langle(\nabla\Theta)^2\rangle_{\mathcal{H},z}$ and its bulk fraction $\epsilon_{\Theta,\text{bulk}}$ by restricting $z\in[\lambda_\Theta,1-\lambda_\Theta]$. The fraction of thermal dissipation in the bulk $\epsilon_{\Theta,\text{bulk}}/\epsilon_\Theta$ (Fig.~\ref{fig:BLkin}(a,b)) grows with increasing $\Ra$ as the flow slowly changes from boundary-layer-dominated to bulk-dominated thermal dissipation in the non-rotating limit \citep{grossmann_scaling_2000,stevens_unifying_2013}. The bulk fraction also grows for rapid rotation in rotation-dominated regime, where suppressed vertical motion causes a stronger vertical temperature gradient. However, in the small-$\Ra$ subregime with vertically coherent vortices, the bulk fraction first decreases with increasing rotation, which indicates that a strongly boundary-layer-dominated thermal dissipation is required for a large enhancement of the heat transport. This reduction vanishes in the large-$\Ra$ subregime, when the vortices start to loose their coherence and leak heat. In there, the bulk fraction roughly increases with $\Ek$. It seems that the truncation of heat transport enhancement giving the maximum location at large $\Ra$ roughly happens around a bulk fraction of $\epsilon_{\Theta,\text{bulk}}/\epsilon_\Theta\approx35\%$. In the following, we lay out what could cause this increase of the bulk fraction towards the geostrophic and rotation-affected turbulence flow state.\\

Starting from a different perspective, we propose that shearing and viscous effects might play a crucial role with increasing $\Ra$ and $\iRo$. For non-rotating RBC, the so-called wind of turbulence that drives the large-scale circulation strengthens with increasing $\Ra$, which leads to larger shearing at the walls. Likewise, faster rotation leads to stronger Ekman vortices and stronger vortical motion in the horizontal plane, especially at the kinetic boundary layer height, which similarly increases the wall shear stress. According to this conceptual idea, we consider the dimensionless wall shear stress as the vertical gradient of the horizontal RMS velocity at the plates $\tau_w=\left.\partial_z u_\mathcal{H}^\text{RMS}\right\vert_\text{wall}$ normalized by the horizontal RMS velocity at boundary layer height $U_V=\left.u_\mathcal{H}^\text{RMS}\right\vert_{\lambda_u}$ as a proxy for the strength of the Ekman vortices (see Supplemental Material \citep{suppmat}, Eqs.~S1,S2 for the full definitions). This ratio yields an inverse length scale, which is also known as the \textit{slope} definition for the kinetic boundary layer \citep{wagner_boundary_2012}:
\begin{equation}\label{eq:BLslope}
\frac{\tau_w}{U_V}=\frac{\left.\partial_z u_\mathcal{H}^\text{RMS}\right\vert_\text{wall}}{\left.u_\mathcal{H}^\text{RMS}\right\vert_{\lambda_u}}\equiv\frac{1}{\lambda_{\tau_w}}
\hspace{0.2cm}\text{.}
\end{equation}
In our context, $\lambda_{\tau_w}$ can be understood as the thickness of a pseudo shear boundary layer.\\

Indeed, we find that the ratio of wall shear stress to vortex strength grows with increasing $\Ra$ and $\iRo$ (Fig.~\ref{fig:BLkin}(c,d)). Moreover, the location of the heat transport maxima follows a fixed contour of $\lambda_{\tau_w}^{-1}$. The global trend of $\lambda_{\tau_w}^{-1}$ is thereby mainly determined by the wall gradient itself (Supplemental Material \citep{suppmat}, Fig.~S1).
In general, all contours of $\lambda_{\tau_w}^{-1}$ follow a constant $\Ek$ in the rotation-dominated regime, which is an indicator for a classical Ekman type scaling for $\lambda_{\tau_w}\propto\Ek^{1/2}$. However, they start to deviate in the rotation-affected regime. Fitting the data for $\iRoopt(\Ra\geq7\cdot10^8)$, we obtain an effective scaling of $\iRoopt\propto\Ra^{-0.77\pm0.06}$ ($\iEkopt\propto\Ra^{-0.27\pm0.06}$) for $\Pr=4.38$ and $\iRoopt\propto\Ra^{-0.87\pm0.06}$ ($\iEkopt\propto\Ra^{-0.37\pm0.06}$) for $\Pr=6.4$. Especially for $\Pr=4.38$, the contours of $\lambda_{\tau_w}^{-1}$ nicely follow the fitted trend $\lambda_{\tau_w}\propto\Ro\,\Ra^{-0.77}$ ($\propto\Ek\,\Ra^{-0.27}$) through the rotation-affected regime. 
We note that due to the decreasing magnitude of enhancement and the broadening peak with increasing $\Ra$, the exact values of $\iRoopt$ show larger uncertainties independent of the good statistical convergence (see App.~\ref{sec:appA}, Tab.~\ref{tab:pr4},\ref{tab:pr6}). This makes the good correlation between the heat transport maxima and $\lambda_{\tau_w}$ even more remarkable.
In the direct comparison of $\lambda_{\tau_w}^{-1}$ (Fig.~\ref{fig:BLkin}(c,d)) with the classical RMS definition of the kinetic boundary layer $\lambda_{u}$ (Fig.~\ref{fig:BLkin}(e,f)), one can clearly observe the significant deviation of $\lambda_{\tau_w}^{-1}$ from a constant $\Ek$, whereas $\lambda_{u}$ shows perfect Ekman type behavior through the entire rotation-dominated and rotation-affected regimes.\\

From a physical point of view, we explain this threshold behavior as follows. On the one hand, an increased shearing at the plates destabilizes the emerging Ekman vortices, which supports their breakup in the bulk. On the other hand, the stronger the vortices are, the faster they rotate at the height of the Ekman layer, and thus the larger $U_V$ becomes. The balance of these two effects is reflected in their ratio $\lambda_{\tau_w}^{-1}$. Therefore, this ratio depicts the relative impact of shearing on the Ekman vortices. When the ratio $\lambda_{\tau_w}^{-1}$ increases (i.e., for a thinner shear boundary layer), the Ekman vortices are perturbed by the leading geostrophic force balance in the bulk, which triggers their heat leakage and their breakup. Both effects would abruptly reduce the heat transport. We again emphasize the broadened enhancement peak at large $\Ra$; a sign for a gradual change of the bulk structure interfering with the enhancing effect of the varying boundary layer ratio. Finally, the optimal interference of the two is resembled by the threshold $\lambda_{\tau_w,\text{crit.}}^{-1}$, beyond which the impeding effects exceed the enhancing effects for the heat transport.\\

The critical threshold $\lambda_{\tau_w,\text{crit.}}^{-1}$ for the cutoff of heat transport enhancement slightly differs with $\Pr$: $\lambda_{\tau_w,\text{crit.}}^{-1}\approx264$ for $\Pr=4.38$ versus $\lambda_{\tau_w,\text{crit.}}^{-1}\approx288$ for $\Pr=6.4$. It shows that the transition is not simply given by $\Ek$ as Ekman-type scaling does not suggest a significant $\Pr$ dependence (Supplemental Material \citep{suppmat}, Fig.~S4). 
A $\Pr$-dependent threshold $\lambda_{\tau_w,\text{crit.}}$ seems reasonable in the context of Ekman vortices getting destabilized by increasing wall shear stress. A larger $\Pr$ means a relatively smaller thermal diffusivity $\kappa$ such that heat diffuses less easily out of the Ekman vortices, while they also remain more stable due to the relatively larger viscosity $\nu$. Thus, stronger shearing is required compared to the vortex strength to break their coherence and terminate heat transport enhancement, resulting in a larger threshold of $\lambda_{\tau_w}^{-1}$ and a shift of the transition to geostrophic turbulence towards larger $\Ra$ and smaller $\Ek$.\\

Last, we provide an attempt to resolve the $\Pr$ dependence and to collapse the optimal rotation rate in the large-$\Ra$ subregime. Therefore, we recall that the transitional Rayleigh numbers suggest a fixed Grashof $\Gr_t=\Ra_t/\Pr\approx10^8$ for the transition between the small-$\Ra$ and large-$\Ra$ subregimes. In the small-$\Ra$ subregime, the optimal rotation rate for varying $\Pr$ collapses on $\iEkopt\propto\Ra^{2/3}$ (see Sec.~\ref{subsec:BLratio}), which will still hold when rewritten to $\Gr$: $\iEkopt\Pr^{-2/3}\propto\Gr^{2/3}$. Note that this factor of $\Pr^{-2/3}$ even remains unchanged when switching from $\iEk$ to $\iRo=\iEk\Gr^{1/2}$: $\iRoopt\Pr^{-2/3}\propto\Gr^{1/6}$. In terms of the scaling theory by \citet{ecke_turbulent_2023}, the $\Pr$ dependence is then given by $\Pr^{-2\gamma}$. Having (i) a collapsed arm of $\iEkopt$ (resp. $\iRoopt$) in the small-$\Ra$ subregime, (ii) a collapsed transition to the large-$\Ra$ sub regime, and (iii) mostly the same scaling on the arm of $\iEkopt$ (resp. $\iRoopt$) in the large-$\Ra$ subregime will lead to a universal behavior of the $\Pr$ compensated optimal rotation $\iEkopt\Pr^{-2/3}$ (resp. $\iRoopt\Pr^{-2/3}$) as a function of $\Gr$ (Fig.~\ref{fig:sumup}(b)). In this way, we can also determine a comprehensive effective scaling for the optimal rotation rate among both $\Pr$ by fitting $\iEkopt\Pr^{-2/3}$ (resp. $\iRoopt\Pr^{-2/3}$) for $\Gr>1.2\cdot10^{8}$, which yields $\iEkopt\Pr^{-2/3}\propto\Gr^{-0.32\pm0.15}$ (resp. $\iRoopt\Pr^{-2/3}\propto\Gr^{-0.82\pm0.15}$). These exponents are one-to-one comparable to those fitted above and -- not surprisingly -- lie in the middle of the separately fitted values. We note that more data are required to improve the fit of these exponents and reduce their uncertainties.

\vspace{-.05cm}
\section{\label{sec:conc}Conclusions}

\begin{figure}
\begin{minipage}[b]{.5\textwidth}
\centering
\includegraphics[width=\textwidth]{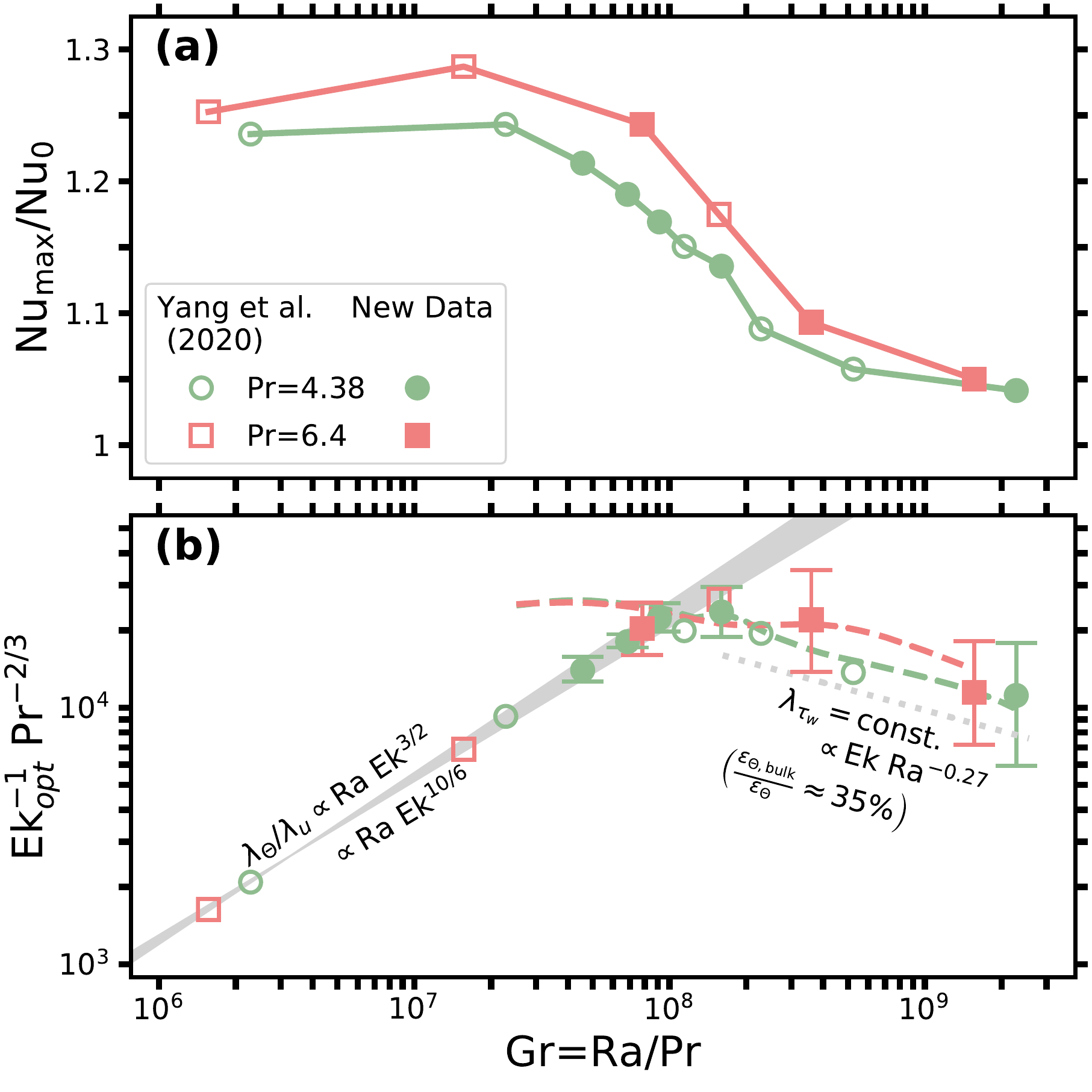}
\end{minipage}
\hspace{.02\textwidth}
\begin{minipage}[b]{.469\textwidth}
\caption{\label{fig:sumup}Summary of findings in this study: \textbf{(a)}~maximal enhancement of the non-rotating heat transport $\Nu_{\mathrm{max}}/\Nu_0$ and \textbf{(b)}~the corresponding optimal rotation rate as compensated inverse Ekman number $\iEkopt\Pr^{2/3}$ as a function of $\Gr=\Ra/\Pr$ (open symbols - data points from \citet{yang_what_2020}, filled symbols - new data points). At small $\Ra$ ($\Gr\lesssim10^8$), $\iEkopt(\Ra)$ follows the universal scaling for a constant boundary layer ratio $\lambda_\Theta/\lambda_u\approx1\propto\Ra\,\Ek^{3/2}$ \citep{king_heat_2012,yang_what_2020} or $\propto\Ra\,\Ek^{10/6}$ \citep{ecke_turbulent_2023}. At large $\Ra$ ($\Gr\gtrsim10^8$), $\iEkopt(\Ra)$ follows a constant ratio of wall shear stress to vortex strength defining a pseudo shear boundary layer thickness $\lambda_{\tau_w}$ with a $\Pr$-dependent critical threshold $\lambda_{\tau_w,\text{crit.}}(\Pr)$, which corresponds to $\approx35\%$ thermal dissipation in the bulk.}
\end{minipage}
\end{figure}

We investigated the differences for heat transport enhancement $\Nu/\Nu_0$ and its optimal rotation rate $\iRoopt$ (or $\iEkopt$) in rotating RBC for Rayleigh numbers $10^{7}\leq\Ra\leq10^{10}$. Therefore, we further analyzed the data of \citet{yang_what_2020} and extended the data set with newly conducted DNSs for Prandtl numbers $\Pr=4.38$ and $\Pr=6.4$ in the periodic domain. Thereby, the optimal rotation rate follows different effective scalings for small and large $\Ra$, showing a rather sharp transition, whereas the enhancement amplitude decreases rather gradually from $\approx30\%$ to $\approx5\%$ with increasing $\Ra$ across this transition (Fig.~\ref{fig:sumup}). We resolve that the transition appears to be $\Pr$-dependent around $\Ra_t\approx4\cdot10^8$ and $\Ra_t\approx7\cdot10^8$ for $\Pr=4.38$ and $\Pr=6.4$, respectively, which suggests a fixed transitional Grashof number $\Gr_t=\Ra_t/\Pr\approx10^8$.\\

In the \textit{small-$Ra$ subregime}, the optimal rotation rate ($\iRoopt\approx0.12\,\Pr^{1/2}\,\Ra^{1/6}\Leftrightarrow\iEkopt\propto\Ra^{2/3}$) corresponds to a constant ratio of thermal and kinetic boundary layer thicknesses $\lambda_\Theta/\lambda_u\approx1\propto\Ra\,\Ek^{3/2}$, which maximizes the efficiency of Ekman pumping in feeding the vertically coherent vortices in terms of the heat transport \citep{king_boundary_2009,stevens_optimal_2010,kunnen_role_2011}. We find that, for the boundary layer ratio itself, this scaling continues in the \textit{large-$Ra$ subregime}. This means that the deviation of the optimal rotation rate from $\iEkopt\propto\Ra^{2/3}$ and the breakdown of heat transport enhancement at large $\Ra$ must have a different origin.\\

Analyzing the flow morphology, we find a transition of the flow from plumelike, vertically coherent vortices to a rather decorrelated bulk, which shows remarkable similarities to geostrophic turbulence and is therefore considered as its rotation-affected analog. Our data suggest that the changing behavior of heat transport enhancement is directly related to this transition of the flow state. Thereby, the breakdown of the coherent vortices nicely depicts why the boundary layer ratio becomes less important, as those Ekman pumping fed vortices cannot bridge a substantial portion of the bulk.\\

Our analysis reveals that the location of the heat transport maximum in the large-$\Ra$ subregime is determined by a constant ratio of rotation self-induced shearing at the plates to the strengths of the vortices, which defines an (inverse) pseudo shear boundary layer thickness $\lambda_{\tau_w}^{-1}$ (also known as the \textit{slope} definition of the kinetic boundary layer \citep{wagner_boundary_2012}). Beyond this critical threshold of $\lambda_{\tau_w}^{-1}$, at which the bulk fraction of the thermal dissipation $\epsilon_{\Theta,\text{bulk}}/\epsilon_\Theta$ exceeds $\approx35\%$, any further enhancement is truncated. The threshold in $\lambda_{\tau_w}^{-1}$ is $\Pr$-dependent and scales roughly as $\lambda_{\tau_w,\text{crit.}}^{-1}\propto f(\Pr)\,\Ro\,\Ra^{-0.77\pm0.06}$, which yields for the optimal rotation rate:
\begin{equation}
\iRoopt\propto f(\Pr) \cdot \Ra^{-0.77\pm0.06}\;\Leftrightarrow\;\iEkopt\propto \tilde{f}(\Pr) \cdot \Ra^{-0.27\pm0.06}
\hspace{0.2cm}\text{.}
\end{equation}
The rather gradual reduction of the enhancement $\Nu_{\mathrm{max}}/\Nu_0$ across the transition from the small-$\Ra$ to the large-$\Ra$ subregime (Fig.~\ref{fig:sumup}(a)) results from (i) the decreasing coherence of Ekman vortices for the optimal boundary layer ratio $\lambda_\Theta/\lambda_u$ towards the threshold $\lambda_{\tau_w,\text{crit.}}$, and (ii) the $\lambda_\Theta/\lambda_u$ becomes less and less beneficial once the $\lambda_{\tau_w,\text{crit.}}$ is reached. Altogether, our data confirms that heat transport enhancement will vanish for $\Ra\gg10^{10}$ and waterlike $\Pr$ numbers.\\

Finally, our data provide evidence that the optimal rotation rate for maximal heat transport enhancement could be universally given by $[\iEkopt\Pr^{-2/3}](\Gr)$ (Fig.~\ref{fig:sumup}(b)) and $[\iRoopt\Pr^{-2/3}](\Gr)$, respectively. In the future, more data for larger values of $\Pr$ are required to verify or disprove that presumed universal collapse. In both cases, this will improve the understanding of the $\Pr$ dependence of the transition between the small-$\Ra$ and the large-$\Ra$ subregimes, the threshold $\lambda_{\tau_w,\text{crit.}}(\Pr)$, and the enhancement $\Nu_{\mathrm{max}}/\Nu_0$. Eliminating their $\Pr$ effects might help to identify the mechanism behind the aforementioned cutoff of heat transport enhancement. Moreover, the $\Pr$ dependence will allow for an estimate of the general relevance of heat transport enhancement at large $\Ra$ for large-$\Pr$ fluids. In the bigger picture \citep{kunnen_geostrophic_2021}, this study can serve as a stepping stone for future studies to further uncover and quantify the transition from rotation-enhanced RBC at small $\Ra$ to rotation-affected RBC at large $\Ra$ without heat transport enhancement.

\begin{acknowledgments}
We thank Yantao Yang for providing us full access to his data, Olga Shishkina for our continuous collaboration (here in particular the nice discussions about the details of her scaling theory), and Dominik Krug for giving us helpful advises for the spectral analysis. We further thank Chris Howland for an ``epiphany triggering'' PoF seminar. This work was funded by the ERC Starting Grant \textit{UltimateRB} No.~804283. We acknowledge the access to several computational resources, all of which were used for this work: PRACE for awarding us access to MareNostrum 4 based in Spain at the Barcelona Computing Center (BSC) under projects 2020225335, 2020235589, and 2021250115, the Gauss Centre for Supercomputing e.V. (www.gauss-centre.eu) for funding this project by providing computing time on the GCS Supercomputer SuperMUC-NG at Leibniz Supercomputing Centre (www.lrz.de), and NWO Science for the use of supercomputer facilities.\\

\centering\textbf{Data availability statement}\\
The data that support the findings of this study are openly available in \textit{4TU.ResearchData} at \url{http://doi.org/10.4121/ce1fb7ef-8bbe-45e8-a72a-09fefc4bb7ab}.
\end{acknowledgments}

\appendix
\section{\label{sec:appA}Overview of new DNSs}

Tables~\ref{tab:pr4} and \ref{tab:pr6} summarize the most relevant details of the conducted numerical simulations.

\begingroup
\squeezetable
\begin{table*}
\caption{\label{tab:pr4}Summary of numerical parameters for the new DNSs at $\Pr=4.38$: inverse Rossby and Ekman numbers $\iRo$ and $\iEk$; $10\,l_c$ requirement for domain size; width-to-height ratio $\Gamma$ of the DNS; number of grid points in vertical and horizontal direction $N_z$, $N_{x,y}$; Nusselt number $\Nu$; 2nd half Nusselt number $\Nu_h$; number of points within the thermal(kinetic) boundary layer $N_{BL}$; crudest vertical resolution of the Kolmogorov scales $\eta_K$ in the bulk $(\Delta z/\eta_K)_\text{max}$; averaging interval of flow time $\Delta t_\text{avg}$; and maximal CFL number CFL$_\text{max}$ controlling the dynamic time stepping together with a maximal time step $\Delta t_\text{max}=0.005$ in all the cases.}
\begin{ruledtabular}
\begin{tabular}{cccccccccccc}
$\iRo$&$\iEk$&$10\,l_c$&$\Gamma$&$N_z$&$N_{x,y}$&$\Nu$&$\Nu_{h}$&$N_{BL}$&$(\Delta z/\eta_K)_\text{max}$&$\Delta t_\text{avg}$&CFL$_\text{max}$\\
\colrule
\multicolumn{11}{l}{$\Ra=2\cdot10^8$:}&1.1 $\downarrow$\\
$0$			& $0$				& -			&$4$	&256&1024&37.85&37.84&11(35)&0.74&1600\\
$0.\bar{3}$	& $2.25\cdot10^3$	& $3.68$ 	&$4$	&256&1024&40.14&40.15&11(24)&0.77&1600\\
$1$			& $6.76\cdot10^3$	& $2.55$	&$3$	&256&768&41.93&41.93&10(20)&0.77&1600\\
$3.\bar{3}$	& $2.25\cdot10^4$	& $1.71$	&$2$	&256&512&45.43&45.40&11(15)&0.77&1600\\
$4$			& $2.70\cdot10^4$	& $1.61$	&$2$	&256&512&45.88&45.84&11(14)&0.76&1600\\
$4.\bar{4}$	& $3.00\cdot10^4$	& $1.55$	&$2$	&256&512&46.23&46.25&11(14)&0.76&1600\\
$5$			& $3.38\cdot10^4$	& $1.49$	&$2$	&256&512&46.38&46.37&11(13)&0.76&1600\\
$5.\bar{5}$	& $3.75\cdot10^4$	& $1.44$	&$2$	&256&512&46.56&46.54&11(13)&0.75&1600\\
$6.25$		& $4.22\cdot10^4$	& $1.38$	&$2$	&256&512&46.52&46.51&11(12)&0.75&1600\\
$7.14$		& $4.83\cdot10^4$	& $1.32$	&$1.5$	&256&384&46.36&46.30&11(12)&0.74&1600\\
$7.94$		& $5.36\cdot10^4$	& $1.28$	&$1.5$	&256&384&45.91&45.96&11(11)&0.73&1600\\
$10$		& $6.76\cdot10^4$	& $1.18$	&$1.5$	&256&384&44.16&44.22&11(11)&0.71&1600\\
\colrule
\multicolumn{11}{l}{$\Ra=3\cdot10^8$:}&0.6 $\downarrow$\\
$0$			& $0$				& - 			&$4$	&288&1152&42.64&42.68&14(41)&0.77&1600\\
$0.\bar{3}$	& $2.76\cdot10^3$	& $3.44$	&$4$	&288&1152&45.21&45.20&13(29)&0.80&1600\\
$1$			& $8.28\cdot10^3$	& $2.38$	&$3$	&288&864&46.96&46.94&13(24)&0.80&1600\\
$3.\bar{3}$	& $2.76\cdot10^4$	& $1.60$	&$2$	&288&576&50.24&50.27&13(18)&0.80&1600\\
$4$			& $3.31\cdot10^4$	& $1.50$	&$2$	&288&576&50.66&50.58&13(17)&0.80&1600\\
$4.\bar{4}$	& $3.68\cdot10^4$	& $1.45$	&$2$	&288&576&50.86&50.82&13(17)&0.80&1600\\
$5$			& $4.14\cdot10^4$	& $1.39$	&$2$	&288&576&51.05&51.02&13(16)&0.79&1600\\
$5.\bar{5}$	& $4.60\cdot10^4$	& $1.35$	&$2$	&288&576&51.25&51.28&13(16)&0.79&1600\\
$6.25$		& $5.17\cdot10^4$	& $1.29$	&$2$	&288&576&51.26&51.21&13(15)&0.79&1600\\
$7.14$		& $5.91\cdot10^4$	& $1.24$	&$1.5$	&288&432&51.10&51.16&13(15)&0.78&1600\\
$7.94$		& $6.57\cdot10^4$	& $1.19$	&$1.5$	&288&432&50.82&50.79&13(14)&0.77&1600\\
$10$		& $8.28\cdot10^4$	& $1.11$	&$1.5$	&288&432&49.36&49.32&13(13)&0.75&1600\\
\colrule
\multicolumn{11}{l}{$\Ra=4\cdot10^8$:}&0.6 $\downarrow$\\
$0$			& $0$				& - 			&$4$	&320&1280&46.45&46.44&15(45)&0.76&1081\\
$0.\bar{3}$	& $3.19\cdot10^3$	& $3.28$	&$4$	&320&1280&49.21&49.21&14(32)&0.80&1600\\
$1$			& $9.56\cdot10^3$	& $2.27$	&$3$	&320&960&50.95&50.96&14(27)&0.80&1600\\
$3.\bar{3}$	& $3.19\cdot10^4$	& $1.52$	&$2$	&320&640&53.94&53.98&14(20)&0.80&1600\\
$4$			& $3.82\cdot10^4$	& $1.43$	&$2$	&320&640&54.19&54.24&14(19)&0.80&1600\\
$5$			& $4.78\cdot10^4$	& $1.33$	&$1.5$	&320&480&54.60&54.69&14(18)&0.79&1600\\
$5.\bar{5}$	& $5.31\cdot10^4$	& $1.28$	&$1.5$	&320&480&54.63&54.61&14(17)&0.79&1600\\
$6.25$		& $5.97\cdot10^4$	& $1.23$	&$1.5$	&320&480&54.68&54.70&14(17)&0.78&1600\\
$7.14$		& $6.83\cdot10^4$	& $1.18$	&$1.5$	&320&480&54.60&54.56&14(16)&0.78&1600\\
$7.94$		& $7.58\cdot10^4$	& $1.14$	&$1.5$	&320&480&54.40&54.32&14(16)&0.77&1600\\
$10$		& $9.56\cdot10^4$	& $1.05$	&$1.5$	&320&480&53.15&53.11&14(15)&0.75&1600\\
\colrule
\multicolumn{11}{l}{$\Ra=7\cdot10^8$:}&1.1 $\downarrow$\\
$0$			& $0$				& - 			&$4$	&384&1536&55.14&55.12&12(46)&0.73&810\\
$0.\bar{3}$	& $4.21\cdot10^3$	& $2.98$	&$3$	&384&1152&58.17&58.19&11(31)&0.76&1600\\
$1$			& $1.26\cdot10^4$	& $2.07$	&$2$	&384&768&59.88&59.84&11(25)&0.76&1600\\
$1.59$		& $2.01\cdot10^4$	& $1.77$	&$2$	&384&768&60.65&60.65&11(22)&0.77&1600\\
$2.5$		& $3.16\cdot10^4$	& $1.52$	&$2$	&384&768&61.69&61.72&11(19)&0.77&1600\\
$3.\bar{3}$	& $4.21\cdot10^4$	& $1.39$	&$1.5$	&384&576&62.17&62.10&11(18)&0.77&1600\\
$4$			& $5.06\cdot10^4$	& $1.30$	&$1.5$	&384&576&62.28&62.29&11(16)&0.77&1600\\
$5$			& $6.32\cdot10^4$	& $1.21$	&$1.5$	&384&576&62.32&62.37&11(15)&0.77&1600\\
$6.25$		& $7.90\cdot10^4$	& $1.12$	&$1.5$	&384&576&61.98&62.02&11(14)&0.76&1600\\
$7.94$		& $1.00\cdot10^5$	& $1.04$	&$1.5$	&384&576&61.75&61.80&12(13)&0.75&1600\\
$10$		& $1.26\cdot10^5$	& $0.96$	&$1.5$	&384&576&60.78&60.85&11(12)&0.73&1600\\
\colrule
\multicolumn{11}{l}{$\Ra=10^{10}$:}&1.1 $\downarrow$\\
$0$			& $0$				& - 			&$3$	&1024&3072&126.38&126.48&23(104)&0.69&419\\
$0.1$		& $4.78\cdot10^3$	& $2.86$	&$3$	&1024&3072&127.64&127.58&23(97)&0.71&427\\
$0.\bar{3}$	& $1.59\cdot10^4$	& $1.92$	&$2$	&1024&2048&130.09&130.04&22(77)&0.71&811\\
$0.625$		& $2.99\cdot10^4$	& $1.55$	&$2$	&1024&2048&131.01&130.96&22(69)&0.71&800\\
$1$			& $4.78\cdot10^4$	& $1.33$	&$1.5$	&1024&1536&130.80&130.89&22(62)&0.71&1378\\
$1.59$		& $7.58\cdot10^4$	& $1.14$	&$1.5$	&1024&1536&129.39&129.36&21(54)&0.71&1360\\
$3.\bar{3}$	& $1.59\cdot10^5$	& $0.89$	&$1$	&1024&1024&125.83&125.93&22(42)&0.71&1600\\
$10$		& $4.78\cdot10^5$	& $0.62$	&$1$	&1024&1024&105.12&105.15&23(28)&0.68&1600\\
$33.\bar{3}$& $1.59\cdot10^6$	& $0.41$	&$0.5$	&1024&512&71.36&71.06&25(18)&0.57&1600\\
\end{tabular}
\end{ruledtabular}
\end{table*}
\endgroup

\begingroup
\squeezetable
\begin{table*} 
\caption{\label{tab:pr6}Summary of numerical parameters for the new DNSs at $\Pr=6.4$: see Tab.~\ref{tab:pr4} for column details.}
\begin{ruledtabular}
\begin{tabular}{cccccccccccc}
$\iRo$&$\iEk$&$10\,l_c$&$\Gamma$&$N_z$&$N_{x,y}$&$\Nu$&$\Nu_h$&$N_{BL}$&$(\Delta z/\eta_K)_\text{max}$&$\Delta t_\text{avg}$&CFL$_\text{max}$\\
\colrule
\multicolumn{11}{l}{$\Ra=5\cdot10^8$:}&1.1 $\downarrow$\\
$0$		& $0$				& $-$	&$4$&416&1620&49.33&49.40&12(50)&0.48&600\\
$4$		& $3.54\cdot10^4$	& $1.47$&$1.5$&416&625&59.77&59.84&11(17)&0.51&600\\
$5$		& $4.42\cdot10^4$	& $1.36$&$1.5$&416&625&60.55&60.45&11(16)&0.51&600\\
$6.25$	& $5.52\cdot10^4$	& $1.27$&$1.5$&416&625&61.18&61.25&11(15)&0.50&600\\
$7.94$	& $7.01\cdot10^4$	& $1.17$&$1.5$&416&625&61.33&61.26&11(13)&0.50&600\\
$10$	& $8.84\cdot10^4$	& $1.08$&$1.5$&416&625&60.80&60.79&11(12)&0.49&600\\
$12.5$	& $1.10\cdot10^5$	& $1.00$&$1.5$&416&625&59.15&58.96&11(11)&0.48&600\\
\colrule
\multicolumn{11}{l}{$\Ra=2.3\cdot10^9$:}&1.1 $\downarrow$\\
$0$		& $0$				& $-$	&$4$&648&2592&78.90&78.82&12(66)&0.51&410\\
$0.4$	& $7.58\cdot10^3$	& $2.45$&$2.5$&648&1620&83.18&83.24&11(41)&0.53&400\\
$0.625$	& $1.18\cdot10^4$	& $2.11$&$2.5$&648&1620&84.03&84.03&11(38)&0.53&400\\
$1$		& $1.90\cdot10^4$	& $1.81$&$2$&648&1296&84.94&84.94&11(33)&0.54&400\\
$1.59$	& $3.01\cdot10^4$	& $1.55$&$2$&648&1296&85.90&85.91&11(29)&0.54&600\\
$2.5$	& $4.74\cdot10^4$	& $1.33$&$1.5$&648&972&87.23&87.20&11(25)&0.54&600\\
$4$		& $7.58\cdot10^4$	& $1.14$&$1.5$&648&972&88.00&87.96&11(21)&0.55&600\\
$6.25$	& $1.18\cdot10^5$	& $0.98$&$1$&648&648&87.41&87.42&11(17)&0.54&600\\
$10$	& $1.90\cdot10^5$	& $0.84$&$1$&648&648&86.38&86.09&11(14)&0.53&600\\
$15.87$	& $3.01\cdot10^5$	& $0.72$&$1$&648&648&83.04&82.82&11(12)&0.51&600\\
$25$	& $4.74\cdot10^5$	& $0.62$&$1$&648&648&71.31&71.41&11(10)&0.47&600\\
\colrule
\multicolumn{11}{l}{$\Ra=10^{10}$:}&1.1 $\downarrow$\\
$0.4$	& $1.58\cdot10^4$	& $1.92$&$2$&1296&2592&129.76&129.82&28(98)&0.46&492\\
$0.625$	& $2.47\cdot10^4$ 	&$1.65$&$2$&1256&2592&130.42&130.43&28(90)&0.46&500\\
$1$		& $3.95\cdot10^4$	& $1.41$&$1.5$&1296&1944&130.73&130.80&28(81)&0.46&600\\
$1.59$	& $6.27\cdot10^4$	& $1.21$&$1.5$&1296&1944&130.54&130.60&28(71)&0.46&600\\
$2.5$	& $9.88\cdot10^4$	& $1.04$&$1.5$&1296&1944&130.38&130.35&28(61)&0.47&600\\
$4$		& $1.58\cdot10^5$	& $0.89$&$1$&1296&1296&129.34&129.26&28(52)&0.47&600\\
$6.25$	& $2.47\cdot10^5$	& $0.77$&$1$&1296&1296&125.08&125.06&28(44)&0.47&600\\
$10$	& $3.95\cdot10^5$	& $0.66$&$1$&1296&1296&118.29&118.43&28(37)&0.46&600\\
\end{tabular}
\end{ruledtabular}
\end{table*}
\endgroup

\clearpage

\bibliography{literature}

\end{document}


\title{Supplemental Material for ``Optimal heat transport in rotating Rayleigh-B\'{e}nard convection at large Rayleigh numbers''}

\author{Robert Hartmann}
\email[]{r.hartmann@utwente.nl}
\affiliation{Physics of Fluids Group and Twente Max Planck Center, J.M. Burgers Center for Fluid Dynamics, University of Twente, 7500 AE Enschede, The Netherlands}

\author{Guru S. Yerragolam}
\affiliation{Physics of Fluids Group and Twente Max Planck Center, J.M. Burgers Center for Fluid Dynamics, University of Twente, 7500 AE Enschede, The Netherlands}

\author{Roberto Verzicco}
\affiliation{Physics of Fluids Group and Twente Max Planck Center, J.M. Burgers Center for Fluid Dynamics, University of Twente, 7500 AE Enschede, The Netherlands}
\affiliation{Dipartimento di Ingegneria Industriale, University of Rome 'Tor Vergata', 00133 Rome, Italy}
\affiliation{Gran Sasso Science Institute, 67100 L'Aquila, Italy}

\author{Detlef Lohse}
\affiliation{Physics of Fluids Group and Twente Max Planck Center, J.M. Burgers Center for Fluid Dynamics, University of Twente, 7500 AE Enschede, The Netherlands}
\affiliation{Max Planck Institute for Dynamics and Self-Organisation, 37077 G\"ottingen, Germany}

\author{Richard J.A.M. Stevens}
\affiliation{Physics of Fluids Group and Twente Max Planck Center, J.M. Burgers Center for Fluid Dynamics, University of Twente, 7500 AE Enschede, The Netherlands}

\date{August 28, 2023}

\begin{abstract}
\hspace*{-0.8cm}Supplemental Material for the main article \cite{hartmann_optimal_2023} at \url{https://doi.org/10.1103/PhysRevFluids.8.083501}
\end{abstract}

\maketitle

\vspace{-0.2cm}

\section{\label{sec:appB}$\lambda_{\tau_w}^{-1}$ as measure for the strength of Ekman vortices}

\begin{figure*}
\centering
\includegraphics[width=\textwidth]{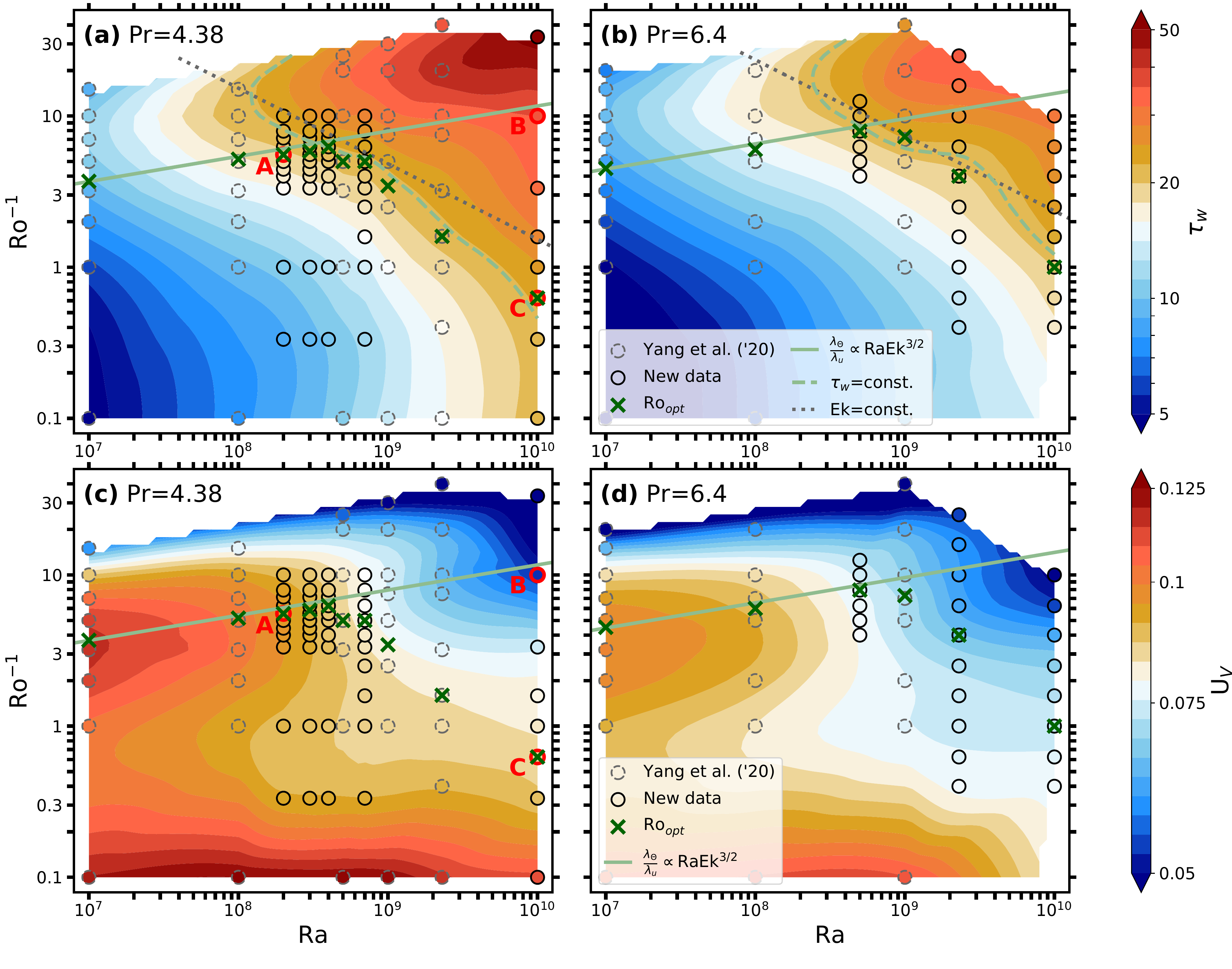}\\
\vspace{-0.3cm}
\caption{\label{fig:wshear}Components of the wall shear stress to vortex strength ratio (inverse \textit{slope} boundary layer $\lambda_{\tau_w}^{-1}=\tau_w/U_V$): \textbf{(a,b)}~wall shear stress $\tau_w=\left.\partial_z u_\mathcal{H}^\text{RMS}\right\vert_\text{wall}$ (Eq.~\ref{eq:wshear}) and \textbf{(c,d)}~maximum horizontal velocity at RMS boundary layer height $U_V=\left.u_\mathcal{H}^\text{RMS}\right\vert_{\lambda_u}$ (Eq.~\ref{eq:uv}) in the parameter space of rotation $\iRo$ and thermal driving $\Ra$ for $\Pr=4.38$ and $\Pr=6.4$, respectively (dashed circles - data points from \citet{yang_what_2020}, closed circles - new data points, background - linear interpolation). The crosses mark the maximal heat transport $\Nu_{\mathrm{max}}(\Ra)$ per $\Ra$ (as in previous figures). The solid line scales as $\lambda_\Theta/\lambda_u\approx1\propto\Ra\,\Ek^{3/2}$ \citep{king_heat_2012,yang_what_2020}. The dashed lines in (a) and (b) emphasize contours of constant $\tau_w$. The dotted lines indicate constant $\Ek$ for comparison. A, B, and C mark the cases presented in Fig.~5~(main article).}
\end{figure*}

In the main article (Sec.~5), we introduced the ratio of the shear stress at the plates $\tau_w$ to the horizontal velocity at kinetic boundary layer height $U_V$ as a proxy for the strength of Ekman vortices. In both, $\tau_w$ and $U_V$, we utilized the RMS of the horizontal velocity, since there is no horizontal mean velocity in rotating RBC. The horizontal RMS velocity is accurately computed from the two lateral velocity components by $u_\mathcal{H}^\text{RMS}=\sqrt{\left< u_x^2 + u_y^2 \right>_\mathcal{H}}$. The average dimensionless wall shear stress at both plates is then computed as:
\begin{equation}\label{eq:wshear}
\tau_w=\left< \left\vert \partial_z u_\mathcal{H} \right\vert \right>_{\mathcal{H},z=\{0,1\}}
\approx \left< \left\vert \partial_z u_\mathcal{H}^\text{RMS} \right\vert \right>_{z=\{0,1\}}
=\left< \left\vert \partial_z \sqrt{\left< u_x^2 + u_y^2 \right>_\mathcal{H}} \right\vert \right>_{z=\{0,1\}}
\hspace{0.2cm}\text{.}
\end{equation}
Under the assumption that horizontal motion is predominantly vortical in rotating RBC, the horizontal RMS velocity would indicate the average tangential velocity at the outer radius of the vortices. Consider that a vortex with constant azimuthal velocity $u_\vartheta$ will cause spatial fluctuations in $u_{x,y}=\pm u_\vartheta$, which are reflected in $u_\mathcal{H}^\text{RMS}$. Therefore, $u_\mathcal{H}^\text{RMS}$ can serve as measure for the average rotation speed of the vortices $U_V$: 
\begin{equation}\label{eq:uv}
U_V=\left< u_\mathcal{H}^\text{RMS} \right>_{z=\{\lambda_u,1-\lambda_u\}}
=\left< \max\limits_{z} \left( u_\mathcal{H}^\text{RMS} \right) \right>_{z\lessgtr0.5}
=\left< \max\limits_{z} \left( \sqrt{\left< u_x^2 + u_y^2 \right>_\mathcal{H}} \right) \right>_{z\lessgtr0.5}
\hspace{0.2cm}\text{.}
\end{equation}
We suppose that the faster a vortex is spinning the more difficult it is to perturb, why $U_V$ can be seen as a proxy for the strength of the vortices. Fig.~\ref{fig:wshear} shows the individual behavior of $\tau_w$ and $U_V$, i.e., the components of the wall shear stress to vortex strength ratio (inverse \textit{slope} boundary layer $\lambda_{\tau_w}^{-1}=\tau_w/U_V$) in the $(\Ra,\iRo)$ parameter space. The main trend of $\lambda_{\tau_w}$ arises from the wall shear stress, which principally increases with increasing $\Ra$ and $\iRo$ (Fig.~\ref{fig:wshear}(a,b)). Further, $\tau_w$ already correlates to some degree with locations of the heat transport maxima at large $\Ra$. Still the normalization by $U_V$ improves the correlation as $\lambda_{\tau_w}$ shows (see main article, Sec.~5, Fig.~9(c,d)). Finally, we note that $U_V$ is largest towards the weakly or non-rotating cases, but also intensifies around the heat transport maxima in the small-$\Ra$ subregime, where vertically-coherent vortices support strong enhancement. This supports the conceptual idea of $U_V$ and $\lambda_{\tau_w}^{-1}$ as measure for the strength of Ekman vortices.

\section{\label{sec:appC}Supplementary Figures}

\begin{figure*}
\centering
\includegraphics[width=\textwidth]{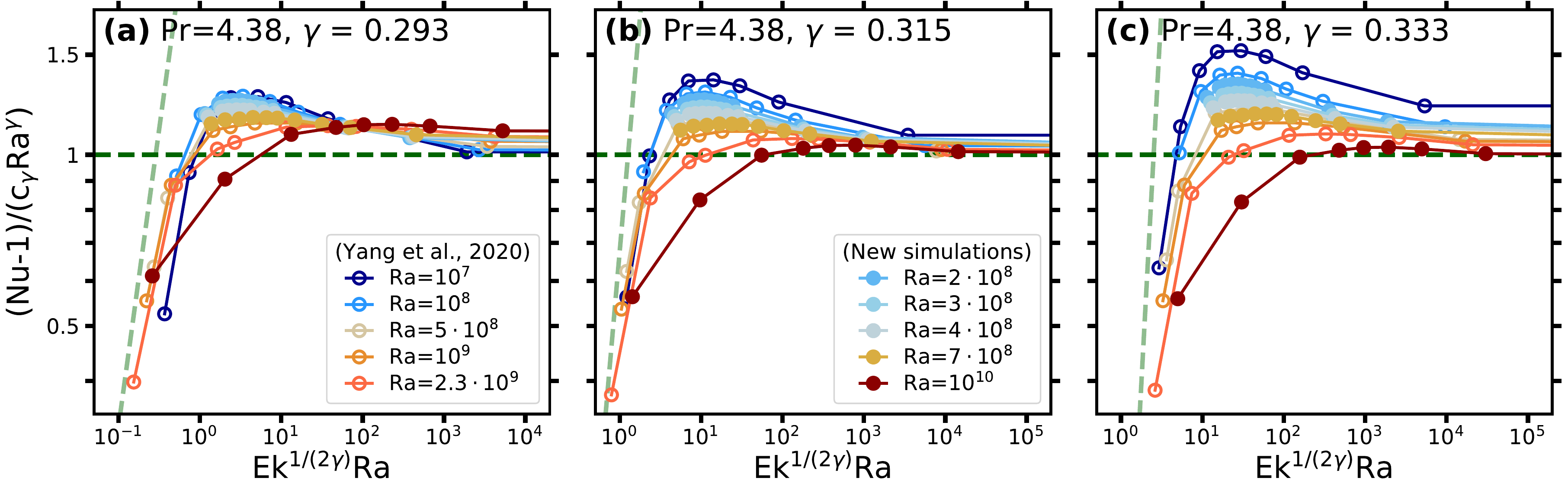}\\
\vspace{-0.46cm}
\includegraphics[width=\textwidth]{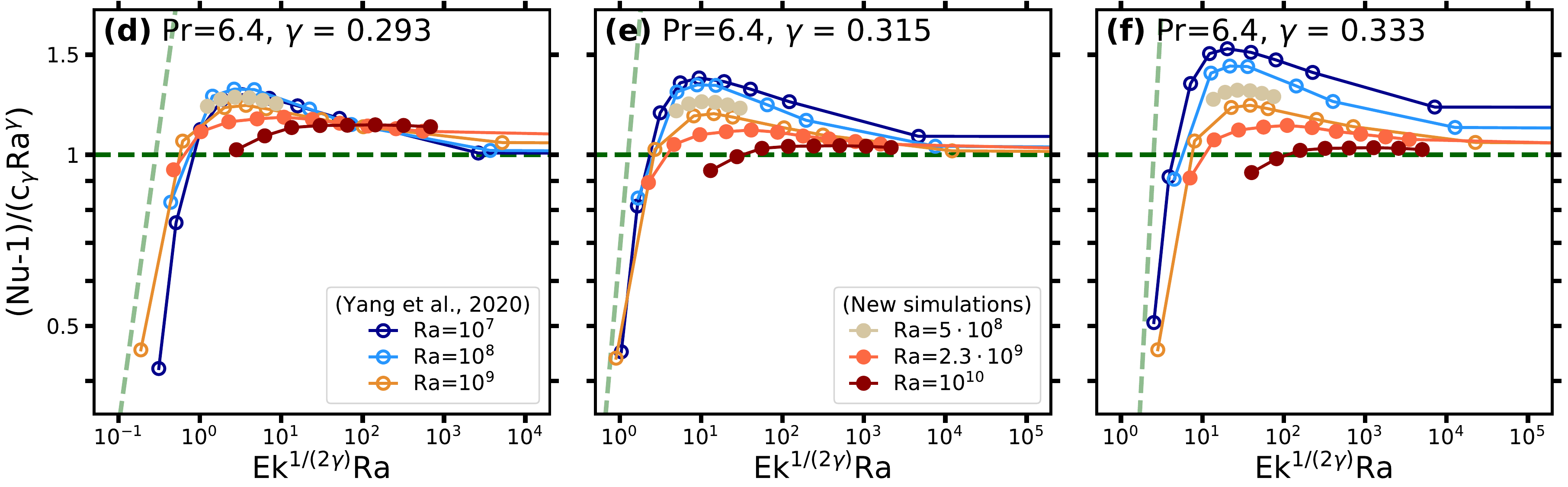}
\caption{\label{fig:othercollapse}Collapse of the heat transport data according to the scaling theory by \citet{ecke_turbulent_2023} for different values of $\gamma$, the exponent of the non-rotating heat transport scaling ($\Nu-1=c_\gamma\Ra^\gamma$) for $\Pr=4.38$~\textbf{(a-c)} and $\Pr=6.4$~\textbf{(d-f)}: \textbf{(a,d)}~$\gamma=0.293$ lower bound of by the small $\Ra$ fit, \textbf{(b,e)}~$\gamma=0.315$ upper bound of the large $\Ra$ fit, and \textbf{(c,f)}~$\gamma=1/3$ theoretical scaling value. The dashed lines indicate the expected scalings in the buoyancy-dominated regime $(\Nu-1)/\Ra^\gamma=const.$ (towards the right) and the rotation-dominated regime $(\Nu-1)/\Ra^\gamma\propto(\Ek^{1/(2\gamma)}\Ra)^{s_{RD}}$ with $s_{RD}=8\gamma^2/(3-8\gamma)$ (towards the left), respectively.}
\end{figure*}

\begin{figure*}
\centering
\includegraphics[width=\textwidth]{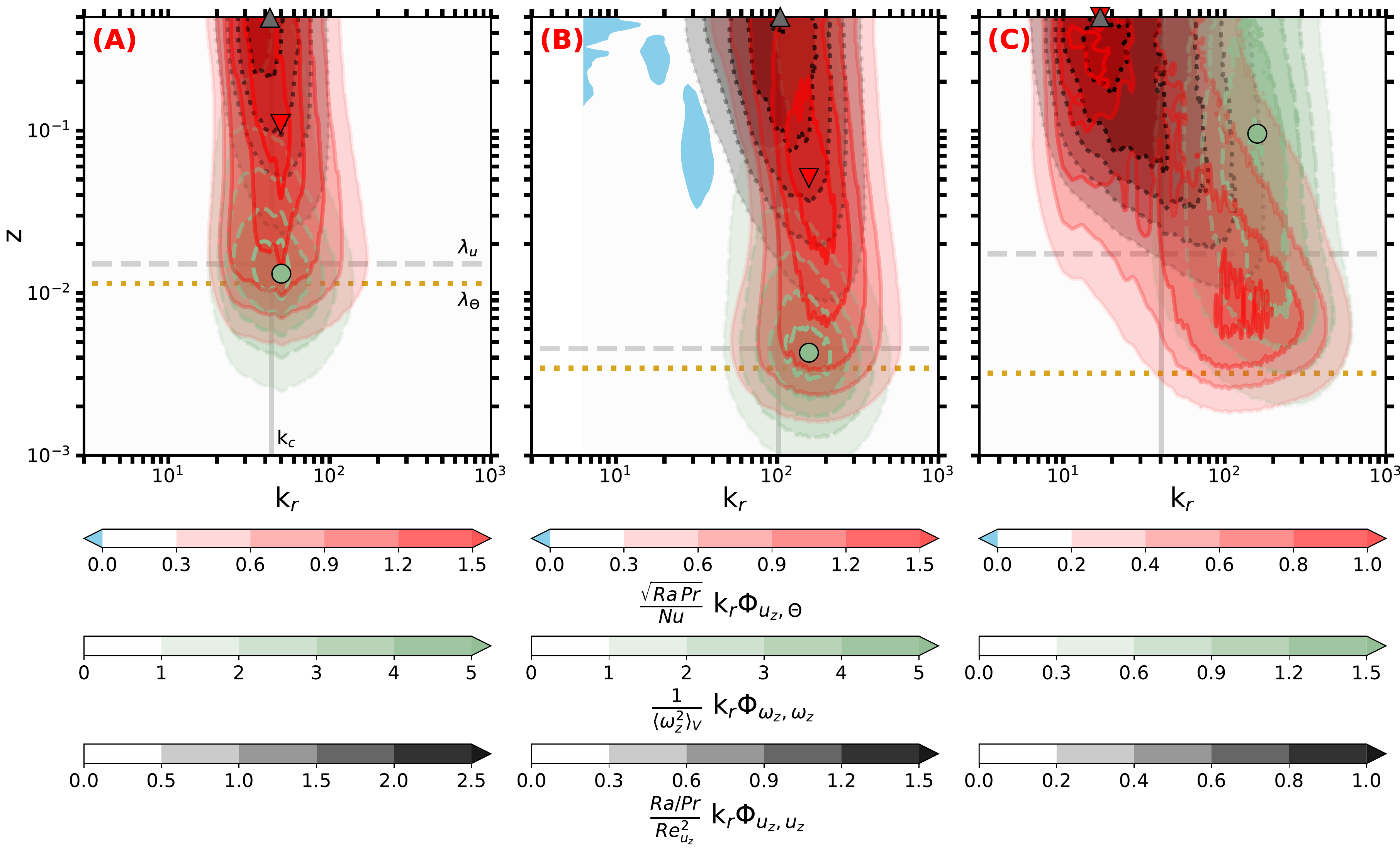}
\caption{\label{fig:kkspectra}Time-averaged, premultiplied radial spectra $k_r \Phi_{q_1,q_2}$ of the vertical velocity ($q_{1,2}=u_z$), the vertical vorticity ($q_{1,2}=\omega_z$), and the local heat transport (cospectra $q_1=u_z, q_2=\Theta$) for the example cases A, B, and C, see Fig.~5 (main article). The upward triangle, the dot, and the downward triangle mark the peak location of $k_r\Phi_{u_z,u_z}$, $k_r\Phi_{\omega_z,\omega_z}$, and $k_r\Phi_{u_z,\Theta}$, respectively. The dashed and dotted lines are kinetic and thermal boundary layer thicknesses. The vertical solid line scales with $k_c=\frac{2\pi}{4.82}\Ek^{-1/3}$ and serves as a guide to the eye for the theoretical size of horizontal structures in rotating RBC.}
\end{figure*}

\begin{figure*}
\centering
\includegraphics[width=.74\textwidth]{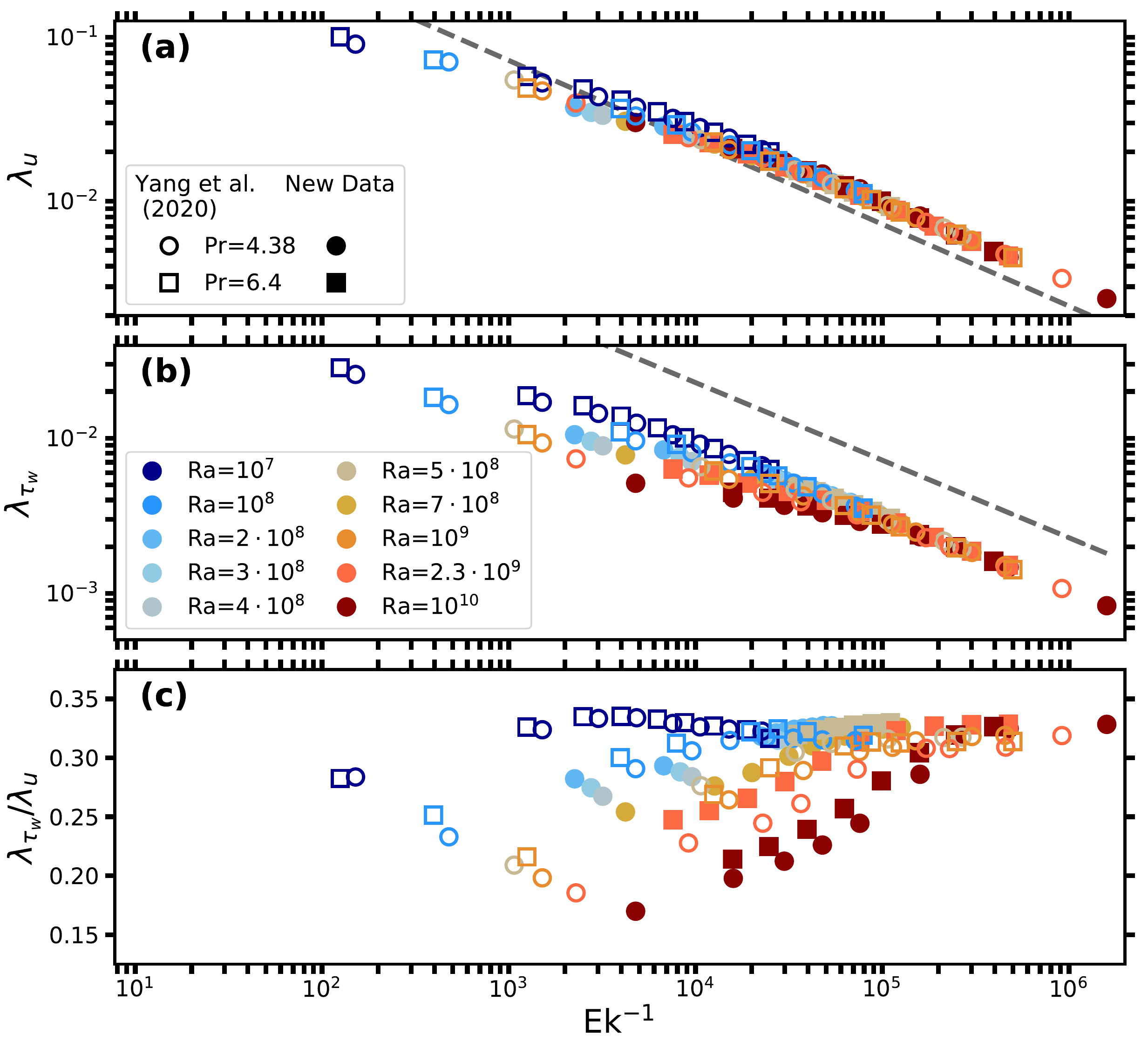}
\caption{\label{fig:BLscale}\textbf{(a)}~Kinetic boundary layer thickness $\lambda_{u}$ (based on the RMS profile), and \textbf{(b)}~pseudo shear boundary layer thickness $\lambda_{\tau_w}$ (based on the \textit{slope} method \citep{wagner_boundary_2012}) as function of the Ekman number $\Ek$. \textbf{(c)}~Ratio of the two boundary layer definitions $\lambda_{\tau_w}/\lambda_{u}$. Open symbols refer to data by \citet{yang_what_2020}, closed symbols to new data. The dashed gray line in (a) and (b) indicates Ekman type scaling $\lambda_u\propto2.284\,\Ek^{1/2}$ \citep[see][]{rajaei_transitions_2016}.}
\end{figure*}

Fig.~\ref{fig:othercollapse} shows the sensitivity of our heat transport data in the scaling theory by \citet{ecke_turbulent_2023} to variations of the exponent $\gamma$ of the underlying non-rotating heat transport scaling ($\Nu-1=c_\gamma\Ra^\gamma$). The chosen values represent a lower and upper bound of $\gamma$ based on the fits including only the data in the small-$\Ra$ and large-$\Ra$ subregimes (Fig.~\ref{fig:othercollapse}(a,d) and \ref{fig:othercollapse}(b,e)), respectively. Finally, we also show data when the theoretical value $\gamma=1/3$ is used (Fig.~\ref{fig:othercollapse}(c,f)).  The good quality of the data collapse, especially for the maximum heat transport, does not significantly change based on the observed variation in $\gamma$ between small and large $\Ra$. Minor changes can only be observed towards the non-rotating and the rapidly-rotating limits, which of course show a better collapse for the those $\Ra$, for which $\gamma$ has been fine-tuned.\newpage

\vspace*{-0.5cm}
Fig.~\ref{fig:kkspectra} shows the premultiplied time-averaged radial spectra $k_r\Phi_{q_1,q_2}$ corresponding to the regular ones $\Phi_{q_1,q_2}$ discussed in the main article (Sec.~4, Fig.~6). For cases A and B, $\Phi_{q_1,q_2}$ and $k_r\Phi_{q_1,q_2}$ do not qualitatively differ (besides a natural shift towards larger $k_r$). For case C, one observes a relative intensification of the large wavenumbers $k_r$ in $k_r\Phi_{q_1,q_2}$ compared to $\Phi_{q_1,q_2}$. On the one hand, this hides the secondary peak in the vorticity spectrum, which aligns with the velocity spectrum and the heat transport cospectrum at smaller $k_r$. On the other hand, it emphasizes the importance of the small-scale vortex structures for the transport of heat from the thermal boundary layer height $\lambda_\Theta$ up to the kinetic boundary layer height $\lambda_u$, above which the larger structures undertake the heat transport.\\

Fig.~\ref{fig:BLscale} shows the (transition to) Ekman-type scaling for the kinetic boundary layer thickness $\lambda_{u}\propto\Ek^{1/2}$ based on the RMS profile and the \textit{slope} method \citep{wagner_boundary_2012}. The RMS definition $\lambda_{u}$ follows Ekman-type scaling for most of the data points (Fig.~\ref{fig:BLscale}(a)), whereas the \textit{slope} definition $\lambda_{\tau_w}$ only approaches Ekman-type scaling at stronger rotation with increasing $\Ra$ (Fig.~\ref{fig:BLscale}(b)). The difference in the transition to Ekman-type scaling, i.e., in the rotation-affected regime, is also reflected in their ratio $\lambda_{\tau_w}/\lambda_{u}$ (Fig.~\ref{fig:BLscale}(c)).\\

\newpage
{~}
\newpage
{~}
\newpage
{~}\vfill

\bibliography{literature}